# Machine learning aided parameter analysis in Perovskite X-ray Detector

Bobo Zhang[1†], Endai Huang[2,3†], Xinyi Du[4,5], Lu Zhang[1], Xiaokang Ma[6], Jiaxue You[7*]

[1]Key Laboratory of Applied Surface and Colloid Chemistry, Ministry of Education; Shaanxi Key Laboratory for Advanced Energy Devices; Shaanxi Engineering Lab for Advanced Energy Technology; Institute for Advanced Energy Materials; School of Materials Science and Engineering, Shaanxi Normal University, Xi'an 710119, China.

[2]Research Institute of Medical and Biological Engineering, Ningbo University, Zhejiang 315211, China

[3]Department of Computer Science and Engineering, The Chinese University of Hong Kong, Hong Kong SAR 999077, China

[4]Department of Chemical and Biomolecular Engineering, National University of Singapore, Singapore, Singapore.

[5]Solar Energy Research Institute of Singapore (SERIS), National University of Singapore, Singapore, Singapore.

[6]State Key Laboratory of Solidification Processing, Northwestern Polytechnical University, Xi'an, Shaanxi 710072, PR China.

[7]Department of Materials Science and Engineering, Hong Kong Institute for Clean Energy City University of Hong Kong, Hong Kong SAR 999077, China

†: BZ and EH contributed equally.

Corresponding authors*: JY: jiaxuyou@cityu.edu.hk

**ABSTRACT:** Many factors in perovskite X-ray detectors, such as crystal lattice and carrier dynamics, determine the final device performance (e.g., sensitivity and detection limit). However, the relationship between these factors remains unknown due to the complexity of the material. In this study, we employ machine learning (ML) to reveal the relationship between 15 intrinsic properties of halide perovskite materials and their device performance (i.e., band gap, carrier mobility-lifetime product, sensitivity, and detection limit). We construct a database of X-ray detectors and train five machine learning models. The results show that the band gap is mainly influenced by the atomic number of the B-site metal, and the lattice length parameter b has the greatest impact on the carrier mobility-lifetime product ($\mu\tau$). An X-ray detector (m-F-PEA)$_2$PbI$_4$ were generated in the experiment and it further verified the accuracy of our ML models. We suggest further study on random forest regression for X-ray detector applications.

**Table of Contents (TOC)**

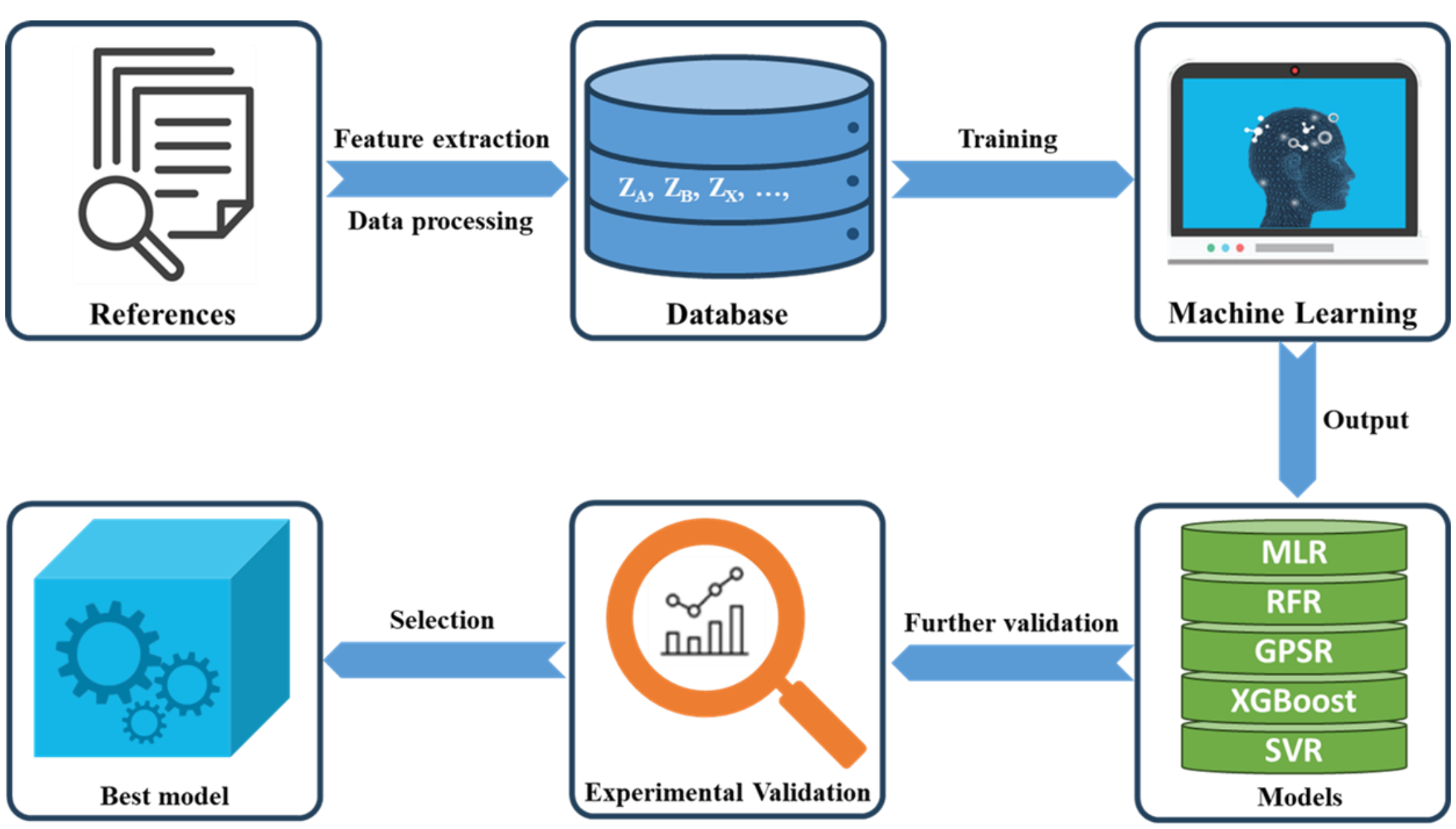

X-ray detectors play a vital role in various fields such as nuclear physics and technology, medical diagnostics, non-destructive testing, security inspections, astronomical observations, and high-energy physics research.[1] Halide perovskites have emerged as a promising candidate for X-ray detection due to their exceptional photoelectric properties, including a large atomic number for high absorption coefficient, a large carrier mobility-lifetime product ($\mu\tau$) for efficient charge collection, and adjustable band gaps leading to low leakage currents.[2, 3] The performance evaluation of X-ray detectors relies heavily on parameters such as sensitivity, detection limit, and dark current.[4] For applications like medical imaging, maintaining exceptional sensitivity and a low detection limit is crucial to minimize radiation exposure. Additionally, these detectors must exhibit dark current densities below 1 nA $cm^{-2}$ to uphold high detection quantum efficiency and dynamic range.[5] Researchers have concentrated on material design, particularly focusing on A-site cations,[6-13] B-site ions,[14-19] and X-site anions regulation.[20, 21]

Utilizing three-dimensional (3D) perovskite materials has enabled X-ray detectors to achieve outstanding performance metrics,[22] although challenges with dark current density persist. Modifying the A-site cations and exploring lower-dimensional structures have shown promise in reducing dark currents.[23] While low-dimensional perovskites exhibit high resistivity and lower dark currents, they may face limitations in carrier transport that impact X-ray responsiveness compared to 3D detectors.[24] Researchers have been striving to enhance the performance and stability of perovskite detectors through various methods, including material synthesis, crystal optimization, and novel fabrication techniques. Despite efforts to improve sensitivity and reduce dark current density, achieving high sensitivity, low detection limits, and minimal dark currents below 1 nA $cm^{-2}$ remains a significant engineering hurdle due to the inherent properties of perovskite materials and technological complexities. In addition, the key parameter affecting dark current remains unknown, due to the complexity of parameter analysis. There are many parameters involved, such as atomic number, lattice constant, trap density (Trap), bandgap ($E_g$), time-resolved photoluminescence (TRPL), resistivity (RT), and carrier mobility-lifetime product ($\mu\tau$) etc.

Machine learning (ML) technology has recently shown promising advancements in the realm of new energy materials design and analysis, particularly in optimizing solar cells, predicting band gaps, designing compositions, and screening lead-free alternatives.[25-28] ML's ability to extract information from complex datasets autonomously can aid in identifying patterns and optimizing key parameters for improved detector performance by linking material properties to detector characteristics affected by multiple factors. Integrating ML methodologies could offer a novel approach to addressing these complex systems and predicting X-ray detector performance.

To this end, this study integrates ML into analyzing X-ray detector performance. The workflow is as follows (Figure 1). Initially, seventy feature parameters ($Z_A$, $Z_B$, $Z_X$, $Z_{total}$, a, b, c, $\alpha$, $\beta$, $\gamma$, $E_g$, $\mu\tau$, TRPL, Trap, RT, detection limit, and sensitivity) are derived from existing literature.. Subsequently, the suggested characteristic parameters are normalized to ensure scale uniformity and enhance convergence speed. Furthermore, five ML algorithms, namely multivariable linear regression (MLR), random forest regression (RFR), eXtreme Gradient Boosting (XGBoost), support vector regression (SVR), and symbolic regression based on genetic programming (GPSR) are then employed to determine the weights of the characteristic parameters that influence $E_g$, $\mu\tau$, sensitivity, and detection limit. Ultimately, a novel X-ray detector device is fabricated to validate the model's accuracy in real-world applications.

The main contributions of this study are summarized as following. First, a database of perovskite X-ray detector was created, containing 137 data samples and 23 features. The database is publicly open-source and can be further used by other researchers. Secondly, five different ML algorithms were applied to analyze the relationship between four target features (i.e., $E_g$, $\mu\tau$, sensitivity, and detection limit) and other parameters showing the applicability of ML into perovskite X-ray detector. Our new experimental material for X-ray detectors further verified the prediction accuracy of ML in the real-world settings. Finally, the analysis of different ML models reveals that the atomic number of B-site elements has the most profound impact on the $E_g$ while lattice length b affects $\mu\tau$ mostly among other features. The data and code are available at https://github.com/hed115599/ML_X-ray_Detector.

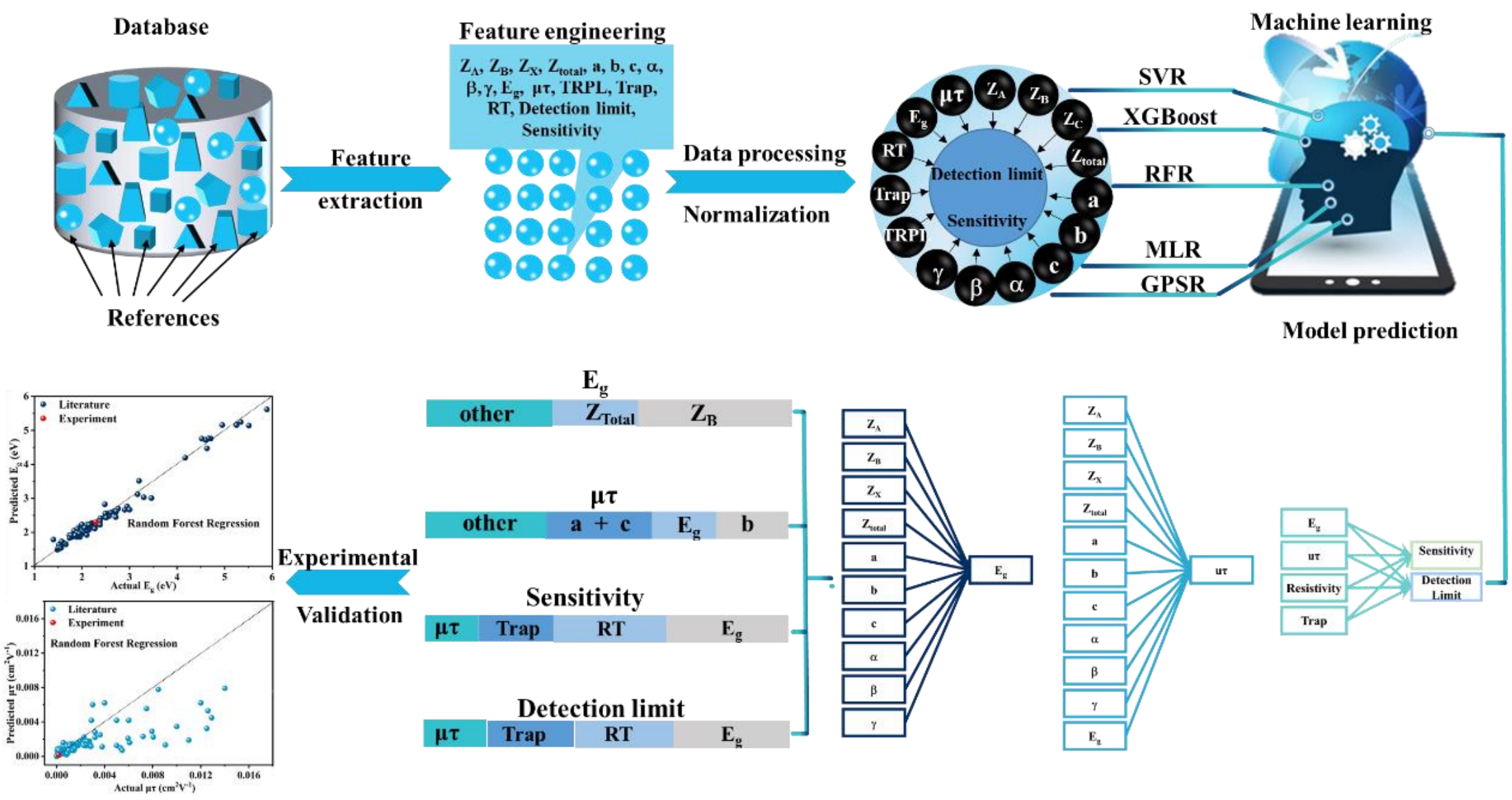

Figure 1. The workflow of this study.

The database is constructed from 136 relevant research papers and 1 real-world experiment conducted by this study. Each data sample contains 23 features, and the details are summarized in Table 1. During the data collection process, missing values (e.g., those not mentioned in the literature) are estimated using comparable values from other literature sources. To mitigate errors introduced by human factors, a diminished weight was applied to data samples containing estimated values when using the data (details provided in the Supplementary Information).

**Table 1. Summary of the database.**

| Feature | Abbreviation | Mean | Max | Min |
| --- | --- | --- | --- | --- |
| **Perovskite** | NA | NA | NA | NA |
| **Molecular Structure** | NA | NA | NA | NA |
| **Atomic number of A site** | $Z_A$ | 109.3 | 1678.0 | 18.0 |
| **Atomic number of B site** | $Z_B$ | 181.2 | 209.0 | 18.0 |
| **Atomic number of X site** | $Z_X$ | 105.0 | 127.0 | 35.5 |
| **Total atomic number** | $Z_{total}$ | 1054.0 | 6443.0 | 210.5 |
| **lattice lengths/a(Å)** | a | 11.0 | 50.7 | 5.9 |
| **lattice lengths/b(Å)** | b | 10.0 | 32.5 | 4.8 |
| **lattice lengths/c(Å)** | c | 13.6 | 57.0 | 5.9 |

| lattice angle/α° | α | 89.4 | 105.1 | 24.9 |
|---|---|---|---|---|
| **lattice angle/β°** | β | 90.4 | 116.8 | 9.6 |
| **lattice angle/γ°** | γ | 94.9 | 120.0 | 9.4 |
| **Band gap (eV)** | $E_g$ | 2.4 | 7.1 | 1.0 |
| **carrier mobility-lifetime product ($cm^2V^{-1}$)** | uτ | 2.1 | 200.0 | 1.1E-06 |
| **Carrier mobility ($cm^2V^{-1}s^{-1}$)** | u | 73.5 | 2652.0 | 2.0E-05 |
| **Resistivity (Ω·cm)** | RT | 7.6E+11 | 6.3E+13 | 1.0E-10 |
| **Trap density ($cm^{-3}$)** | Trap | 3.3E+14 | 2.5E+16 | 1.5E-11 |
| **TRPL(ns)** | TRPL | 4805.6 | 6.0E+05 | 0.5 |
| **Type** | NA | NA | NA | NA |
| **X-ray energy(KeV)** | NA | 57.1 | 140.0 | 8.0 |
| **Applied electric filed (V $mm^{-1}$)** | NA | 357.6 | 5000.0 | 0.0 |
| **Sensitivity (μC $Gy^{-1}$ $cm^{-2}$)** | NA | 1.4E+05 | 5.2E+06 | 0.4 |
| **Detection limit (μ$Gy_{air}^{-1}$ $s^{-1}$)** | NA | 25.0 | 1200.0 | 2.0E-05 |

To demonstrate the machine learning prediction, the Predicted vs. Actual Plot with training and test $R^2$ (RFR as an example) is shown in Figure 2. The x-axis represents the true values while the y-axis represents the predicted values output by the ML model. A closer alignment of points to the diagonal indicates a stronger correspondence between the model's predicted values and the actual values, reflecting the model's predictive accuracy. The training $R^2$ and test $R^2$ of the RFR model for $E_g$ (Figure 2a) are 0.93 and 0.80, respectively. The scatter points locate closely to the diagonal. These demonstrate that the RFR model fits and predicts $E_g$ well. However, for μτ, sensitivity, and detection limit, both the training $R^2$ and test $R^2$ are relatively low, and the Predicted vs. Actual Plot shows errors (Figure 2b-d). Other ML models also show similar results (Table S1). Considering the low training $R^2$ and the substantial learning capabilities of these models, the current feature set may not adequately capture the essential predictors required for forecasting three target objectives. Additionally, the database used in this study was extracted from various literature sources. This aggregation of data from different experimental settings and instrumentation may introduce variability and errors that could further contribute to the observed low predictive accuracy. Further investigations into feature selection, as well as the exploration of additional features, may be necessary to enhance the model's predictive accuracy and these remain to be our future work.

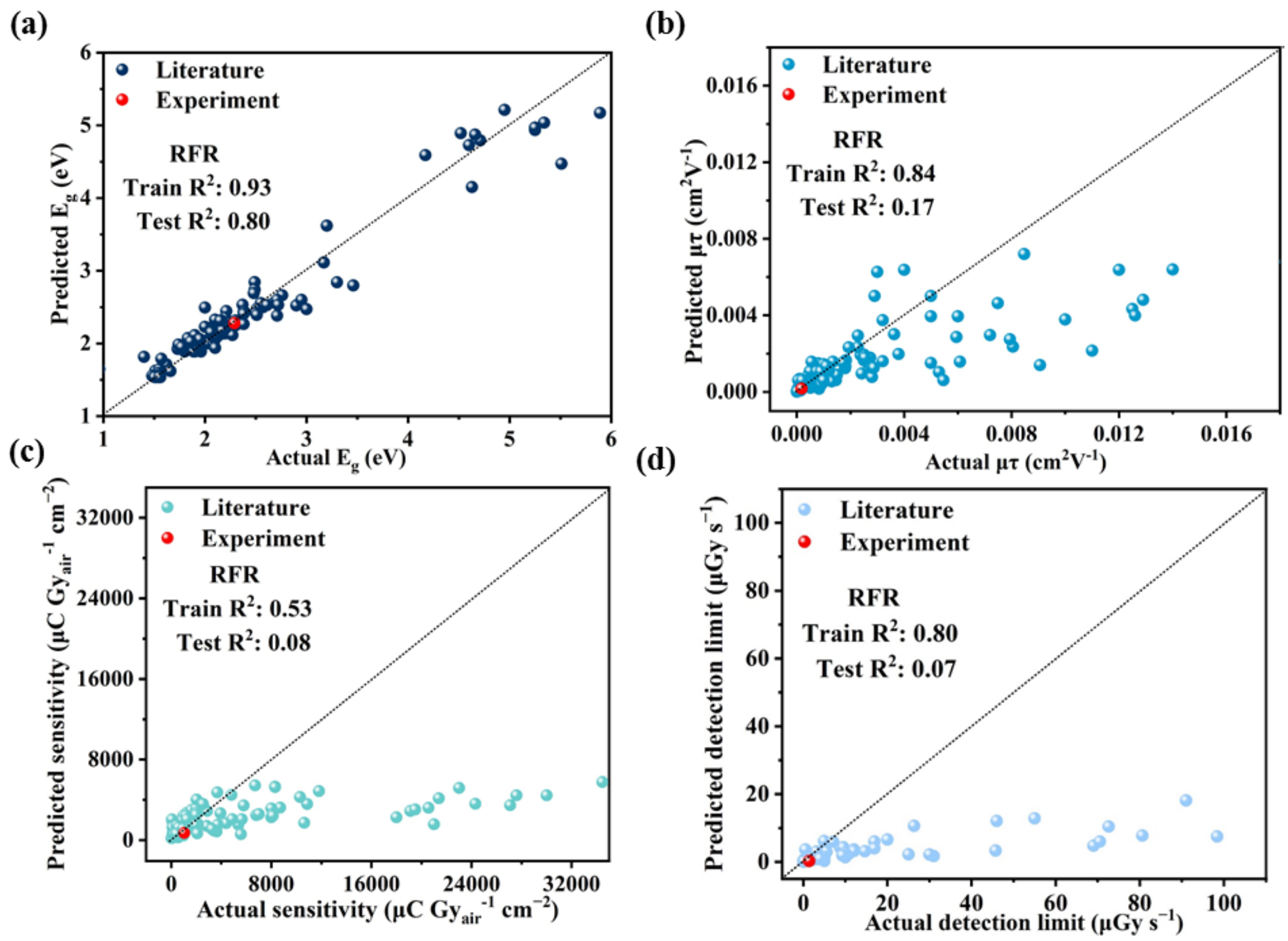

Figure 2. Predicted vs. Actual Plot of RFR for (a) $E_g$, (b) μτ, (c) sensitivity, and (d) detection limit.

To further test the validity of our trained ML models in the real-world instead of only in literature, an experimental verification was conducted. In our database, the majority of samples have μτ values ranging from $1\times10^{-4}$ to $8\times10^{-3}$ $cm^2$ $V^{-1}$, sensitivity ranges from 581 to 5000 μC $Gy_{air}^{-1}$ $cm^{-2}$, and detection limits range from 0.85 to 20 μ$Gy_{air}$ $s^{-1}$. Additionally, it is noteworthy that 70% of the database is comprised of lead-based perovskite materials. Consequently, we have opted to employ a low-dimensional lead-based perovskite material for experimental validation.

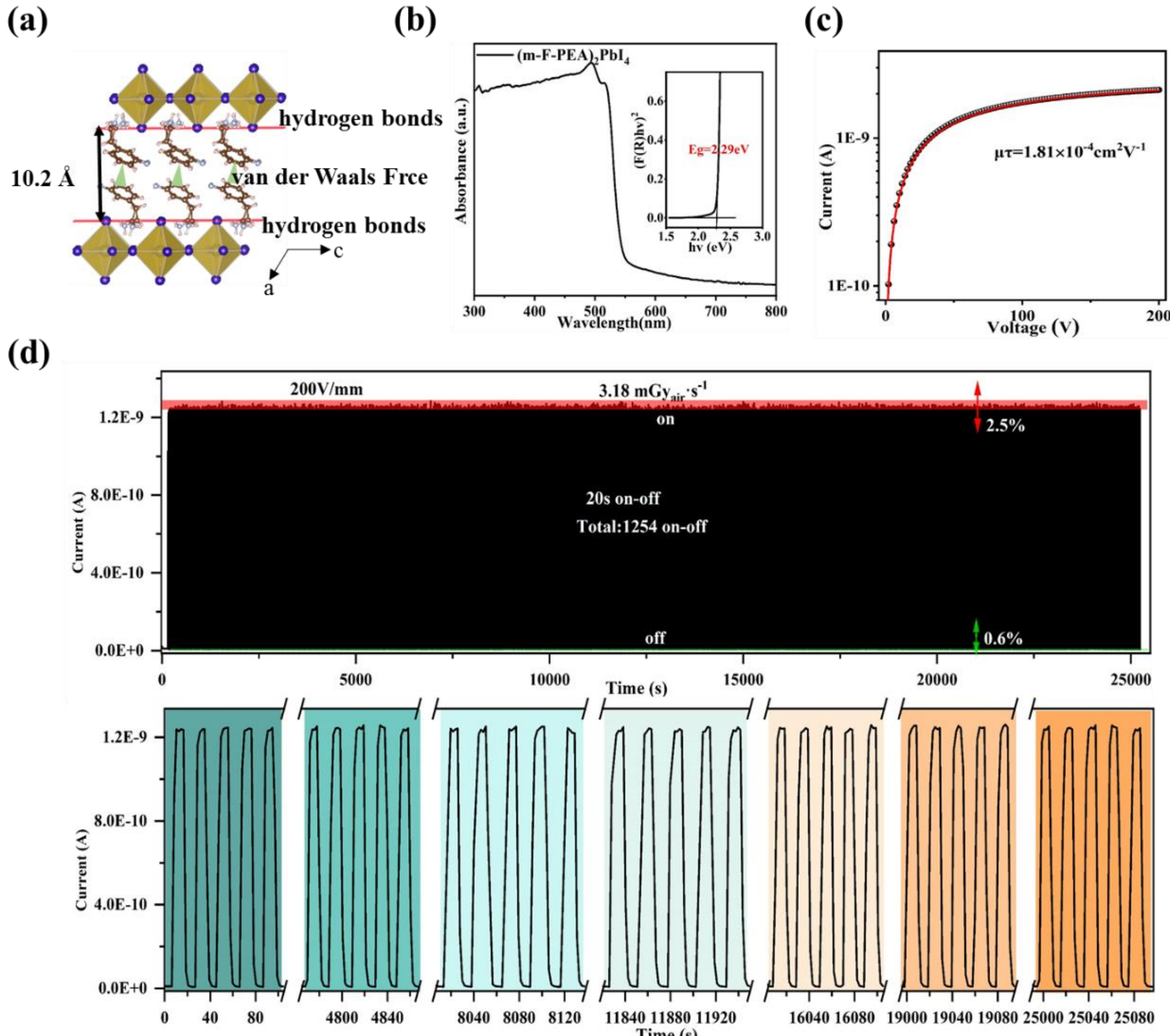

Figure 3. (a) Packing view along b-axis of $(m\text{-}F\text{-}PEA)_2PbI_4$, red represents hydrogen bonding between organic and inorganic layers, and cyan represents the existence of van der Waals forces between organic ions. (b) Absorption spectra for the $(m\text{-}F\text{-}PEA)_2PbI_4$. (c) Photoconductivity of the $(m\text{-}F\text{-}PEA)_2PbI_4$ SC device. (d) $(m\text{-}F\text{-}PEA)_2PbI_4$ SC detector switch stability test and 7 local magnification images.

The block-shaped orange-yellow single crystals $(m\text{-}F\text{-}PEA)_2PbI_4$ (i.e., 20× 5× 0.7 $mm^3$) were grown through temperature cooling crystallization in a heated HI solution containing chemically reactive m-FPEA and PbO (Figure S1a). The results obtained from the scanning electron microscope (SEM) indicate that the surface of the single crystal is exceptionally smooth (Figure S1b), whereas the cross-sectional image (Figure S1c) reveals a periodic layered morphology. The results of the thermogravimetric analysis (TGA) and corresponding first derivative results indicated that the mass loss of $(m\text{-}F\text{-}PEA)_2PbI_4$ was as low as 1% at a temperature of up to 240 °C (Figure S2a,b), which surpasses many hybrid perovskite counterparts, such as $(PEA)_2PbI_4$ (215 °C,

PEA= phenethylammonium)[19], DABCO-$NH_4Cl$ (208.9 °C, DABCO =N-N′-diazabicyclo[2.2.2]octonium).[20] The simulated patterns align closely with the powder X-ray diffraction (XRD) patterns of (m-F-PEA)$_2$Pb$I_4$ (Figure S2c). The periodic XRD peaks observed in the (m-F-PEA)$_2$Pb$I_4$ SCs confirm the presence of a layered crystal structure and its growth along the (002) crystal face orientation (Figure S2d). The precise crystal structure of (m-F-PEA)$_2$Pb$I_4$ was determined by single crystal X-ray diffraction and detailed crystallographic data are provided in Table S2. The crystallographic analysis reveals that the crystal structure of (m-F-PEA)$_2$Pb$I_4$ is a typical Ruddlesden−Popper (RP) phase, adopting the C2/c space group (a=66.489(4) Å, b=12.2142(6) Å, c=12.2131(8) Å). The inorganic layers are separated by two large organic cations, with van der Waals forces acting between the organic layers. The organic layers are primarily connected to the inorganic layers through hydrogen bonding (Figure 3a).

The electronic structure of perovskites is primarily determined by the Pb$I_6$ octahedron that make up the inorganic layer. Consequently, any distortion in the octahedra would affect their optoelectronic properties, such as the absorption/emission of excitons.[29] Octahedral distortion can be quantitatively evaluated using the distortion index (D) and the bond angle variance ($\sigma^2$):

$$\mathrm{D} = \frac{1}{6}\sum_{1}^{6}\frac{|L_i - l_{av}|}{L_{av}} \tag{1}$$

$$\sigma^2 = \sum_{1}^{12}\frac{(\theta_i - 90)^2}{11} \tag{2}$$

The variables $L_\mathrm{i}$ and $L_\mathrm{av}$ represent the individual and average Pb−I bond length, respectively, while $\theta_\mathrm{i}$ denotes the individual I−Pb−I angle. The presence of smaller σ (1.25 deg$^2$) and D (0.01) values in (m-F-PEA)$_2$Pb$I_4$ indicates a narrower fluorescent emission spectrum (Figure S3a).[30] The band gap is acknowledged to be influenced by the Pb−I−Pb bond angles, whereby smaller angles contribute to a reduced overlap between the Pb 6s and I 5p orbitals, consequently yielding a higher band gap.[31] The compounds (m-F-PEA)$_2$Pb$I_4$ (151.1°) and (o-F-PEA)$_2$Pb$I_4$ (150.9°)[32] exhibit comparable Pb-I-Pb bond angles, resulting in a nearly identical band gap (2.29 eV) (Figure S3b and Figure 3b). The band gap (2.10 eV) obtained from density functional theory (DFT) calculations is in close agreement with the experimental results (Figure S4a). Furthermore, the smaller out-of-plane distortion of single-layer perovskites (1.39°) results in narrow-spectrum fluorescence with small Stokes shifts (Figure S3c and Figure S4b). A shorter fluorescence lifetime ($\tau_{ave}$=0.65 ns) indicates that (m-F-PEA)$_2$PbI$_4$ is a typical RP-type 2D perovskite (Figure S4c).

For semiconductor detectors, the trap density and the bulk resistivity of the semiconductor material significantly affect the sensitivity and detection limit of the detector. This is particularly important for X-ray detectors that operate under high electric fields. A low trap density is advantageous for improving the sensitivity of the detector, while a high bulk resistivity is beneficial for enhancing the signal-to-noise ratio of the detector and consequently reducing the detection limit. We initially employed an impedance analyzer to test the capacitance-frequency curve of the (m-F-PEA)$_2$PbI$_4$ SC devices in the frequency range of 0.4 to 1 MHz, in a dark environment, for the purpose of calculating the relative dielectric constant ($\varepsilon$) of the single crystal (Figure S5a). $\varepsilon$ of the single crystal was then calculated to be 6.0 ± 0.4 (Figure S5b). Subsequently, the I-V curve of the (m-F-PEA)$_2$PbI$_4$ SC under dark conditions was analyzed using the Keysight B2912A precision source meter, and evaluated based on the SCLC model (Figure S5c). The defect density of (m-F-PEA)$_2$PbI$_4$ SC is $6.4 \times 10^{10}$ cm$^{-3}$, comparable to that of previously reported 3D[33] and 2D[7] metal halide SCs and significantly lower than traditional semiconductor materials such as Si[34] and CdTe.[35] Additionally, the carrier mobility ($9.76\times10^{-2}$ cm$^2$V$^{-1}$s$^{-1}$) of (m-F-PEA)$_2$PbI$_4$ SC was obtained by analyzing the "child" segment of the dark I-V curve. As shown in Figure S5d, the (m-F-PEA)$_2$PbI$_4$ SC demonstrates a significant bulk resistivity of $3.78 \times 10^{11}$ $\Omega\cdot$cm, which serves to effectively suppress leakage current and mitigate the impact of noise current.

In order to investigate the X-ray detection capabilities of (m-F-PEA)$_2$PbI$_4$, the X-ray absorption coefficients for perovskite and conventional semiconductor materials in the photon energy range of 10 keV to 1.1 MeV were calculated using the NIST X-COM database.[36] The absorption coefficient of (m-F-PEA)$_2$PbI$_4$ is significantly higher than silicon across the entire energy spectrum of the display, rivaling $\alpha$-Se and other similar 2D perovskite materials (Figure S6a). On this basis, we further calculate the thickness required for different materials to fully absorb a fixed energy of X-rays. As shown in Figure S6b, it can be observed that a single crystal of $(m\text{-}F\text{-}PEA)_2PbI_4$ with a thickness of 0.7 mm can absorb 94% of the peak energy of X-rays at 40 keV. Furthermore, based on the modified Hecht equation, the current-voltage (I-V) curve of the $(m\text{-}F\text{-}PEA)_2PbI_4$ SC device under X-ray irradiation was fitted, yielding a mobility-lifetime ($\mu\tau$) product of $1.81\times10^{-4}$ $cm^2$ $V^{-1}$, which is comparable to that of 3D $MAPbBr_3$ ($2.6\times10^{-4}$ $cm^2$ $V^{-1}$)[37] and significantly higher than that of α-Se ($10^{-7}$ $cm^2$ $V^{-1}$)[38] (Figure 3c).

Through the utilization of the exceptional charge transfer properties within the plane of two-dimensional perovskites, a coplanar detector was fabricated based on the $(m\text{-}F\text{-}PEA)_2PbI_4$ SC, and its X-ray detection performance was evaluated using sensitivity and detection limit. Figure S7a shows the on/off photocurrent response curve of the $(m\text{-}F\text{-}PEA)_2PbI_4$ SC X-ray detector at bias voltages from 10 V to 100 V and at X-ray doses from 2.7 μGy $s^{-1}$ to 22.7 μGy $s^{-1}$. The sensitivity at different bias voltages was calculated from the photocurrent on/off response curves as shown in Figure S7b, and reached 1033.7 μC $Gy_{air}^{-1}$ $cm^{-2}$ at a bias voltage of 100 V. The sensitivity of this value is higher than that of a 2D perovskite $(PEA)_2PbI_4$ SC X-ray detector (848 μC $Gy_{air}^{-1}$ $cm^{-2}$)[19], and 51.7 times higher than the sensitivity of the current commercially available α-Se X-ray detector used for X-ray imaging under high electric field (10000 V $mm^{-1}$) (20 μC $Gy_{air}^{-1}$ $cm^{-2}$).[38]Moreover, the response current of the $(m\text{-}F\text{-}PEA)_2PbI_4$ SC detector at low X-ray dose rates was tested for the calculation of the detection limit (Figure S7c). As illustrated in Figure S7d, we calculated and fitted the X-ray dose rate versus detector signal-to-noise ratio (SNR) and extended the fit line to an SNR of 3 (as defined by the International Union of Pure and Applied Chemistry[39]) to determine the detection limit of 1.41 μGy $s^{-1}$ for the $(m\text{-}F\text{-}PEA)_2PbI_4$ SC detector, which is nearly 4 times lower than the dose rate used in conventional medical diagnostics (5.5 μGy $s^{-1}$).[40] The lower detection limit can significantly reduce the radiation dose received by patients during routine X-ray examinations, thereby greatly reducing the risk of X-ray exposure.

The stability and repeatability of detectors in different environments are crucial indicators for evaluating their usability in practical applications. Initially, under the continuous application of an electric field of 200 V $mm^{-1}$, we conducted a dark current test on $(m\text{-}F\text{-}PEA)_2PbI_4$ SC devices for nearly 4 h (Figure S8a). The dark current drift of the 2D $(m\text{-}F\text{-}PEA)_2PbI_4$ SC detector is 9.59 $\times 10^{-8}$ nA $cm^{-1}$ $s^{-1}$ $V^{-1}$, which is significantly lower by five orders of magnitude compared to the dark current drift of the 3D $MAPbI_3$ SC detector (2.0 $\times 10^{-3}$ nA $cm^{-1}$ $s^{-1}$ $V^{-1}$).[41] This remarkable improvement is attributed to the extremely low ion migration of the 2D $(m\text{-}F\text{-}PEA)_2PbI_4$ SC. Furthermore, the $(m\text{-}F\text{-}PEA)_2PbI_4$ SC detector can operate for more than 12 h under high doses and continuous X-ray irradiation, with no significant decrease in X-ray response current, indicating that indicating is intrinsically structurally stable under X-ray irradiation. We also conducted tests on the stability of the $(m\text{-}F\text{-}PEA)_2PbI_4$ SC X-ray detector when exposed to air at room temperature (Figure S8b). The results indicate that the response current of the detector remains comparable to the initial response value, even under different high doses (96.2-790.2 μGy $s^{-1}$) after 60 days (Figure S8c). Finally, after 1254 response cycles at 20 seconds interval, the response current and dark current of the $(m\text{-}F\text{-}PEA)_2PbI_4$ SC detector remained 97.5% and 99.45% of the initial value, respectively (Figure 3d). During a testing period of over 25000 s , the on-off response of seven magnified sections remained consistent, further indicating the device has excellent on-off stability. The response time of detectors has been investigated due to the rapid response of X-rays, which can reduce the radiation dose received by patients in medical diagnosis. The rise time (from 10% to 90%) of the $(m\text{-}F\text{-}PEA)_2PbI_4$ SC detector for pulsed X-ray is measured to be 16.8 ms, while the fall time (from 90% to 10%) is determined to be 29.1 ms (Figure S9). Combining good X-ray detection properties with excellent stability, $(m\text{-}F\text{-}PEA)_2PbI_4$ SC hold promise for high-quality X-ray imaging.

The experimental result above is visualized in Figure 2 (red dot). The RFR model predicted a bandgap value as 2.23 eV, which is closest to the experimental value (2.29 eV), followed by XGBoost (2.16 eV), GPSR (1.99 eV), SVR (2.59 eV), and MLR (1.96 eV) (Figure S10a). Compared to the DFT calculation result (2.1 eV), the prediction by machine learning models is more accurate. Despite the low fitting accuracy of $\mu\tau$ , the RFR model still predicted with considerable accuracy (Figure S10b) for $\mu\tau$. For sensitivity and detection limit, there is a slight error between the experimental value and the predicted value (Figure S10c,d).

To determine which feature has the most significant impact on the target, we ranked the feature importance (calculated using permutation importance) within each ML model (Figure 4 and Figure S11). The ranking reveals that the $Z_B$ has most significant effect on $E_g$ with a highest feature importance among other features. This finding is consistent with the high Pearson correlation coefficient (-0.85) between $E_g$ and $Z_B$ (Figure 4a). The B-sites occupying the perovskites in the collected database are dominated by the elements Pb, Bi and small organic molecules ($NH_4^+$, $N_2H_4^+$). Perovskites incorporating small organic molecules at the B-site are classified as molecular perovskites.[42] The organic constituents of these molecular perovskites play a significant role in determining the band gap. Typically, organic molecules exhibit a wide band, and the weak interaction between organic macromolecules and halogens leads to a relatively large energy discrepancy (i.e., band gap) between the valence and conduction bands.[14, 21, 43]

The GPSR model can provide a formula for predicting bandgaps that

$$E_g = \log(Z_B \times Z_x) + \sqrt{Z_B} \quad (3)$$

where $Z_B$ and $Z_X$ represent the atomic numbers of the B-site and X-site atoms in organic-inorganic hybrid halide perovskites. The formula also indicates the high relationship between $Z_B$ and $E_g$. According to formula 3, the bandgap of perovskites is related to the B and X site atoms but not to the A atom, aligning with the fact that A-site ions do not directly contribute to the band structure in actual perovskites.[44]

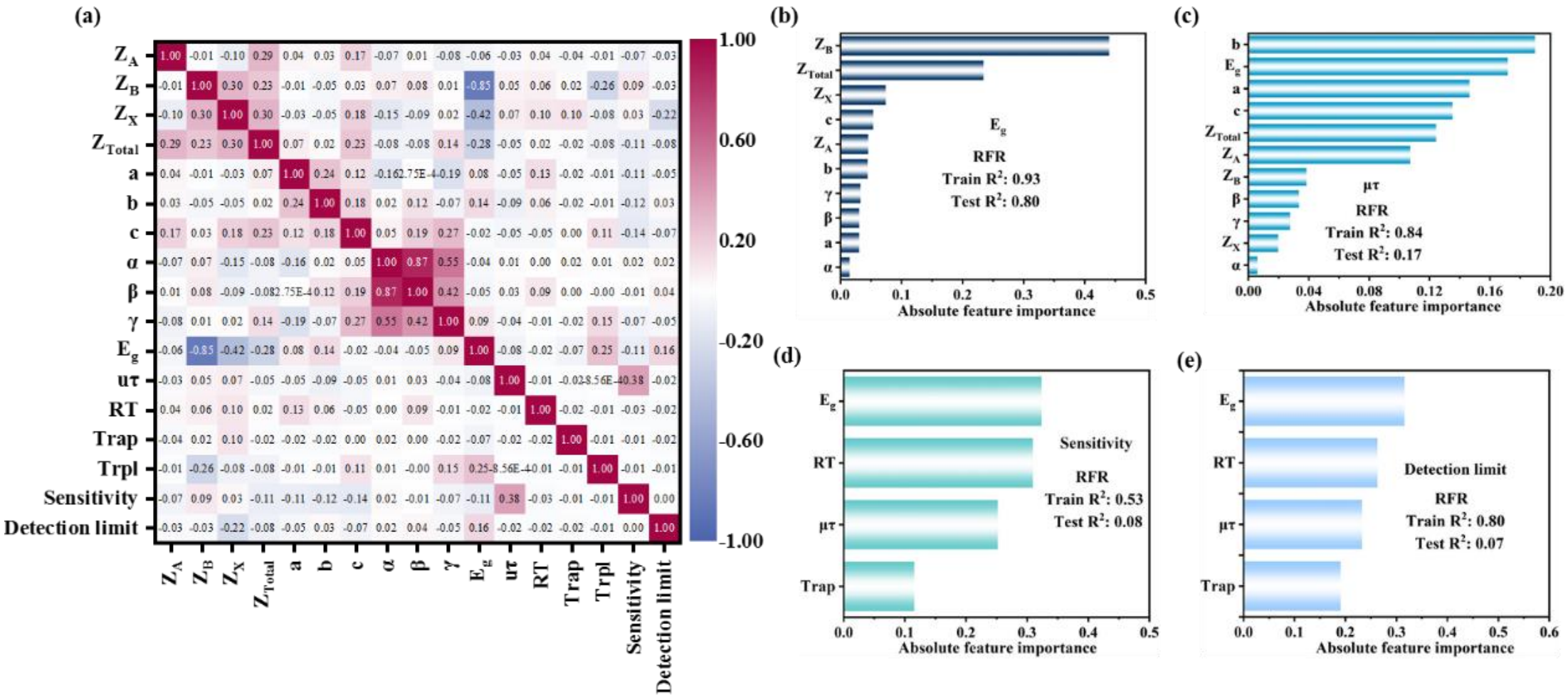

Figure 4. The Pearson correlation heatmap (a) and feature importance ranking using RFR for $E_g$ (b), μτ (c), sensitivity (d), and detection limit (e).

Despite the low test $R^2$ for μτ, most of ML models weight more on lattice length b in prediction of μτ (Figure 4c and Figure S12). Changing the lattice length b can alter the band gap width and band curvature of a crystal, directly impacting the effective mass of charge carriers. A smaller effective mass of charge carriers usually results in higher mobility.[13] Additionally, variations in lattice length b may affect the defect density of the crystal, including vacancies, impurities, dislocations, and more. An increase in defect density can enhance carrier scattering, reduce mobility, and potentially act as non-radiative recombination centers, thereby shortening the carrier lifetime.[45, 46] Due to the low test $R^2$ for sensitivity and detection limit (Figure 4d,e and Figure S13,14), it is hard to draw a conclusion about whether $E_g$ effect these two targets mostly among the four features, which requires further verification.

Among the five ML models, RFR has demonstrated a remarkable ability to predict the properties of perovskite X-ray detectors with high accuracy (Table S1). This robust performance can be attributed to several inherent characteristics of the RFR algorithm and the specific nature of perovskite materials and their properties. Firstly, perovskite materials exhibit complex and non-linear relationships between their structural parameters. RFR is particularly adept at capturing these non-linear interactions without the need for explicit model specification, unlike linear regression models. This capability allows RFR to model the nuanced impacts of compositional and structural variations in perovskites more effectively. Secondly, our datasets may contain outliers or missing values due to experimental challenges and variability in synthesis conditions. RFR is capable of handling such datasets effectively, as the random forest algorithm is less sensitive to outliers and can handle missing data by using surrogate splits or ignoring missing values during the node splitting in trees. Therefore, given these strengths, we suggest further utilization and investigation of RFR for researchers and developers in the field of X-ray detector.

In this work, we collected a database for perovskite X-ray detectors and demonstrated that ML techniques can be applied to analyze parameters in perovskite X-ray detectors. Five ML regression models were trained to analyze the main factors affecting $E_g$, $\mu\tau$, sensitivity, and detection limit. The RFR model has identified the intrinsic parameter $Z_B$ of a given perovskite material to have the greatest impact on the $E_g$, while lattice length b affects ut mostly among the features. The obtained result aligns with its corresponding physical significance, revealing the applicability of the ML model. Our ultimately prepared $(m\text{-}F\text{-}PEA)_2PbI_4$ single-crystal X-ray detectors futher verified the accuracy of ML models in real-world applications. We suggest further studies on RFR for X-ray detector applications.

## ASSOCIATED CONTENT

### Supporting Information

The supporting information is free of charge via the Internet at http://pubs.acs.org.

Detailed description of the machine learning algorithm; materials, growth of single crystals (SCs), fabrication of SC X-ray detectors; material characterization; band structure calculation, dielectric measurement, SCLC measurement, photoconductivity analysis, dark current drift measurements, X-ray detector performance measurements (PDF)

Dataset:

https://view.officeapps.live.com/op/view.aspx?src=https%3A%2F%2Fraw.githubusercontent.com%2Fjoshua-you%2FPerovskite-X-ray-Detector-machine-learning%2Frefs%2Fheads%2Fmain%2Fdataset-2025.10.29.xlsx&wdOrigin=BROWSELINK

**Author contributions**

†BZ and EH contributed equally to this paper. J.Y. conceived the project and wrote the original manuscript. J.Y. supervised the project. B.Z. conducted most of experiments and co-wrote the paper. E.H. performed machine learning analysis. X.D. assisted in testing the X-ray detection data. All authors discussed and analyzed the data.

**Notes**

The authors declare no competing financial interest.

**ACKNOWLEDGEMENTS**

This work was funded by the Strategic Priority Research Program of the Chinese Academy of Sciences (Grant No. XDA17040506), the National Key Research and Development Program of China (2017YFA0204800), the Key Program project of the National Natural Science Foundation of China (51933010), the National Natural Science Foundation of China (61974085/51901190), the 111 Project (B21005), the Young Scientist Initiative Project of School of Materials Science and Engineering at Shaanxi Normal University (SNNU) (2022YSIP-MSE-SNNU001), the Changjiang Scholars and Innovative Research Team (IRT14R33) and the National University Research Fund (GK202103113).

## REFERENCES

(1) He, X.; Deng, Y.; Ouyang, D.; Zhang, N.; Wang, J.; Murthy, A. A.; Spanopoulos, I.; Islam, S. M.; Tu, Q.; Xing, G.; Li, Y.; Dravid, V. P.; Zhai, T. Recent Development of Halide Perovskite Materials and Devices for Ionizing Radiation Detection. *Chem Rev.* **2023** *123*, 1207-1261.

(2) Yakunin, S.; Sytnyk, M.; Kriegner, D.; Shrestha, S.; Richter, M.; Matt, G. J.; Azimi, H.; Brabec, C. J.; Stangl, J.; Kovalenko, M. V.; Heiss, W. Detection of X-ray Photons by Solution-processed organic-inorganic Perovskites. *Nat. Photonics* **2015**, *9*, 444-449.

(3) Guan, Q.; You, S.; Zhu, Z. K.; Li, R.; Ye, H.; Zhang, C.; Li, H.; Ji, C.; Liu, X.; Luo, J. Three-Dimensional Polar Perovskites for Highly Sensitive Self-Driven X-Ray Detection. *Angew. Chem. In.t Ed.* **2024**, e202320180.

(4) Wu, H.; Ge, Y.; Niu, G.; Tang, J. Metal Halide Perovskites for X-Ray Detection and Imaging. *Matter* **2021**, *4*, 144-163.

(5) Jin, P.; Tang, Y.; Li, D.; Wang, Y.; Ran, P.; Zhou, C.; Yuan, Y.; Zhu, W.; Liu, T.; Liang, K.; Kuang, C.; Liu, X.; Zhu, B.; Yang, Y. M. Realizing Nearly-zero Dark Current and Ultrahigh Signal-to-noise Ratio Perovskite X-ray Detector and Image Array by Dark-current-shunting Strategy *Nat. Commun.* **2023**, *14*, 626.

(6) Li, H.; Song, J.; Pan, W.; Xu, D.; Zhu, W. A.; Wei, H.; Yang, B. Sensitive and Stable 2D Perovskite Single-Crystal X-ray Detectors Enabled by a Supramolecular Anchor. *Adv. Mater.* **2020**, *32*, e2003790.

(7) Xia, M.; Yuan, J. H.; Niu, G.; Du, X.; Yin, L.; Pan, W.; Luo, J.; Li, Z.; Zhao, H.; Xue, K. H.; Miao, X.; Tang, J. Unveiling the Structural Descriptor of $A_3B_2X_9$ Perovskite Derivatives toward X-Ray Detectors with Low Detection Limit and High Stability. *Adv. Funct. Mater.* **2020**, *30*, 1910648.

(8) Zhang, M.; Xin, D.; Dong, S.; Zhao, W.; Tie, S.; Cai, B.; Ma, Q.; Chen, Q.; Zhang, W. H.; Zheng, X. Methylamine-Assisted Preparation of Ruddlesden-Popper Perovskites for Stable Detection and Imaging of X-Rays. *Adv. Opt. Mater.* **2022**, 2201548.

(9) Shen, Y.; Liu, Y.; Ye, H.; Zheng, Y.; Wei, Q.; Xia, Y.; Chen, Y.; Zhao, K.; Huang, W.; Liu, S. Centimeter-Sized Single Crystal of Two-Dimensional Halide Perovskites Incorporating Straight-Chain Symmetric Diammonium Ion for X-Ray Detection. *Angew. Chem. Int. Ed.* **2020**, *59*, 14896-14902.

(10) Ma, C.; Gao, L.; Xu, Z.; Li, X.; Song, X.; Liu, Y.; Yang, T.; Li, H.; Du, Y.; Zhao, G.; Liu, X.; Kanatzidis, M. G.; Liu, S. F.; Zhao, K. Centimeter-Sized 2D Perovskitoid Single Crystals for Efficient X-ray Photoresponsivity. *Chem. Mater.* **2022**, *34*, 1699-1709.

(11) Wang, Y.; Zhang, S.; Wang, Y.; Yan, J.; Yao, X.; Xu, M.; Lei, X. W.; Lin, G.; Yue, C. Y. 0D triiodide hybrid halide perovskite for X-ray detection. *Chem Commun (Camb)* **2023**, *59*, 9239-9242.

(12) Xu, Y.; Hu, J.; Xiao, X.; He, H.; Tong, G.; Chen, J.; He, Y. Evaporation crystallization of zero-dimensional guanidinium bismuth iodide perovskite single crystal for X-ray detection. *Inorg. Chem. Front.* **2022**, *9*, 494-500.

(13) Yang, X.; Huang, Y.-H.; Wang, X.-D.; Li, W.-G.; Kuang, D.-B. A-Site Diamine Cation Anchoring Enables Efficient Charge Transfer and Suppressed Ion Migration in Bi-Based Hybrid Perovskite Single Crystals. *Angew. Chem. Int. Ed.* **2022**, *61*, 202204663.

(14) Song, X.; Cohen, H.; Yin, J.; Li, H.; Wang, J.; Yuan, Y.; Huang, R.; Cui, Q.; Ma, C.; Liu, S. F.; Hodes, G.; Zhao, K. Low Dimensional, Metal-Free, Hydrazinium Halide Perovskite-Related Single Crystals and Their Use as X-Ray Detectors. *Small* **2023**, *19*, e2300892.

(15) Guo, J.; Xu, Y.; Yang, W.; Xiao, B.; Sun, Q.; Zhang, X.; Zhang, B.; Zhu, M.; Jie, W. High-

Stability Flexible X-ray Detectors Based on Lead-Free Halide Perovskite Cs(2)TeI(6) Films. *ACS Appl. Mater. Interfaces* **2021**, *13*, 23928-23935.
(16) Dong, K.; Zhou, H.; Shao, W.; Gao, Z.; Yao, F.; Xiao, M.; Li, J.; Liu, Y.; Wang, S.; Zhou, S.; Cui, H.; Qin, M.; Lu, X.; Tao, C.; Ke, W.; Fang, G. Perovskite-like Silver Halide Single-Crystal Microbelt Enables Ultrasensitive Flexible X-ray Detectors. *ACS Nano* **2023**, *17*, 1495-1504.
(17) Fan, Q.; Xu, H.; You, S.; Ma, Y.; Liu, Y.; Guo, W.; Hu, X.; Wang, B.; Gao, C.; Liu, W.; Luo, J.; Sun, Z. Centimeter-Sized Single Crystals of Dion-Jacobson Phase Lead-Free Double Perovskite for Efficient X-ray Detection. *Small* **2023**, *19*, e2301594.
(18) Wu, J.; You, S.; Yu, P.; Guan, Q.; Zhu, Z.-K.; Li, Z.; Qu, C.; Zhong, H.; Li, L.; Luo, J. Chirality Inducing Polar Photovoltage in a 2D Lead-Free Double Perovskite toward Self-Powered X-ray Detection. *ACS Energy Lett.* **2023**, *8*, 2809-2816.
(19) Zhang, B.; Xu, Z.; Ma, C.; Li, H.; Liu, Y.; Gao, L.; Zhang, J.; You, J.; Liu, S. First-Principles Calculation Design for 2D Perovskite to Suppress Ion Migration for High-Performance X-ray Detection. *Adv. Funct. Mater.* **2022**, *32*, 2110392.
(20) Cui, Q.; Song, X.; Liu, Y.; Xu, Z.; Ye, H.; Yang, Z.; Zhao, K.; Liu, S. Halide-modulated self-assembly of metal-free perovskite single crystals for bio-friendly X-ray detection. *Matter* **2021**, *4*, 2490-2507.
(21) Li, Z.; Li, Z.; Peng, G.; Shi, C.; Wang, H.; Ding, S. Y.; Wang, Q.; Liu, Z.; Jin, Z. PF(6) (-) Pseudohalides Anion based Metal-Free Perovskite Single Crystal for Stable X-ray Detector to Attain Record Sensitivity. *Adv. Mater.* **2023**, *35*, e2300480.
(22) Song, Y.; Li, L.; Hao, M.; Bi, W.; Wang, A.; Kang, Y.; Li, H.; Li, X.; Fang, Y.; Yang, D.; Dong, Q. Elimination of Interfacial-electrochemical-reaction-induced Polarization in Perovskite Single Crystals for Ultra-sensitive and Stable X-ray Detector Arrays. *Adv. Mater.* **2021**, *33*, e2103078.
(23) Shen, Y.; Ran, C.; Dong, X.; Wu, Z.; Huang, W. Dimensionality Engineering of Organic-Inorganic Halide Perovskites for Next-Generation X-Ray Detector. *Small* **2023**, e2308242.
(24) He, Y.; Hadar, I.; Kanatzidis, M. G. Detecting Ionizing Radiation Using Halide Perovskite Semiconductors Processed through Solution and Alternative Methods. *Nat. Photonics* **2021**, *16*, 14-26.
(25) Lu, S.; Zhou, Q.; Ouyang, Y.; Guo, Y.; Li, Q.; Wang, J. Accelerated Discovery of Stable Lead-free Hybrid Organic-inorganic Perovskites via Machine Learning. *Nat. Commun.* **2018**, *9*, 3405.
(26) Yu, Y.; Tan, X.; Ning, S.; Wu, Y. Machine Learning for Understanding Compatibility of Organic–Inorganic Hybrid Perovskites with Post-Treatment Amines-SVM. *ACS Energy Lett.* **2019**, *4*, 397-404.
(27) Hu, Y.; Hu, X.; Zhang, L.; Zheng, T.; You, J.; Jia, B.; Ma, Y.; Du, X.; Zhang, L.; Wang, J.; Che, B.; Chen, T.; Liu, S. Machine-Learning Modeling for Ultra-Stable High-Efficiency Perovskite Solar Cells. *Adv. Energy Mater.* **2022**, 2201463.
(28) Zhi, C.; Wang, S.; Sun, S.; Li, C.; Li, Z.; Wan, Z.; Wang, H.; Li, Z.; Liu, Z. Machine-Learning-Assisted Screening of Interface Passivation Materials for Perovskite Solar Cells. *ACS Energy Lett.* **2023**, *8*, 1424-1433.
(29) Du, K. Z.; Tu, Q.; Zhang, X.; Han, Q.; Liu, J.; Zauscher, S.; Mitzi, D. B. Two-Dimensional Lead(II) Halide-Based Hybrid Perovskites Templated by Acene Alkylamines: Crystal Structures, Optical Properties, and Piezoelectricity. *Inorg. Chem.* **2017**, *56*, 9291-9302.
(30) Cortecchia, D.; Neutzner, S.; Srimath Kandada, A. R.; Mosconi, E.; Meggiolaro, D.; De Angelis, F.; Soci, C.; Petrozza, A. Broadband Emission in Two-Dimensional Hybrid Perovskites: The Role of Structural Deformation. *J. Am. Chem. Soc.* **2017**, *139*, 39-42.
(31) Gao, L.; Li, X.; Traore, B.; Zhang, Y.; Fang, J.; Han, Y.; Even, J.; Katan, C.; Zhao, K.; Liu, S.; Kanatzidis, M. G. m-Phenylenediammonium as a New Spacer for Dion-Jacobson Two-Dimensional Perovskites. *J. Am. Chem. Soc.* **2021**, *143*, 12063-12073.
(32) Zhang, B.; Zheng, T.; You, J.; Ma, C.; Liu, Y.; Zhang, L.; Xi, J.; Dong, G.; Liu, M.; Liu, S. Electron‐Phonon Coupling Suppression by Enhanced Lattice Rigidity in 2D Perovskite Single Crystals for High‐Performance X‐Ray Detection. *Adv. Mater.* **2023**, *35*, 2208875.
(33) Shi, R.; Pi, J.; Chu, D.; Jia, B.; Zhao, Z.; Hao, J.; Zhang, X.; Dong, X.; Liang, Y.; Zhang, Y.; Liu, Y.; Liu, S. Promoting Band Splitting through Symmetry Breaking in Inorganic Halide Perovskite Single Crystals for High-Sensitivity X-ray Detection. *ACS Energy Lett.* **2023**, *8*, 4836-4847.
(34) Liu, X.; Zhang, H.; Zhang, B.; Dong, J.; Jie, W.; Xu, Y. Charge Transport Behavior in Solution-Grown Methylammonium Lead Tribromide Perovskite Single Crystal Using α Particles. *J. Phys. Chem. C* **2018**, *122*, 14355-14361.
(35) Balcioglu, A.; Ahrenkiel, R. K.; Hasoon, F. Deep-level Impurities in CdTe/CdS Thin-film Solar Cells. *J. Appl. Phys.* **2000**, *88*, 7175-7178.
(36) M.J. Berger, J. H. H., S.M. Seltzer, J. Chang, J.S. Coursey, R. Sukumar, D.S. Zucker, and K. Olsen. *XCOM: photon cross sections database, National Institute of Standards and Technology (NIST).* http://www.nist.gov/pml/data/xcom/index.cfm (accessed 2023.08.01).
(37) Li, L.; Liu, X.; Zhang, H.; Zhang, B.; Jie, W.; Sellin, P. J.; Hu, C.; Zeng, G.; Xu, Y. Enhanced X-ray Sensitivity of $MAPbBr_3$ Detector by Tailoring the Interface-States Density. *ACS Appl. Mater. Interfaces* **2019**, *11*, 7522-7528.
(38) Kasap, S. O. X-ray Sensitivity of Photoconductors: Application to Stabilized α-Se. *J. Phys. D: Appl. Phys* **2000**, *33*, 2853-2865.
(39) Thompson, M.; Ellison, S. L. R.; Wood, R. Harmonized guidelines for single-laboratory validation of methods of analysis - (IUPAC technical report). *Pure Appl. Chem.* **2002**, *74*, 835-855.
(40) Shearer, D. R.; Bopaiah, M. Dose Rate Limitations of Integrating Survey Meters for Diagnostic X-ray Surveys. *Health Phys.* **2000**, *79*, S20-S21.
(41) Liu, Y.; Xu, Z.; Yang, Z.; Zhang, Y.; Cui, J.; He, Y.; Ye, H.; Zhao, K.; Sun, H.; Lu, R.; Liu, M.; Kanatzidis, M. G.; Liu, S. Inch-Size 0D-Structured Lead-Free Perovskite Single Crystals for Highly Sensitive Stable X-Ray Imaging. *Matter* **2020**, *3*, 180-196.
(42) Song, X.; Cui, Q.; Liu, Y.; Xu, Z.; Cohen, H.; Ma, C.; Fan, Y.; Zhang, Y.; Ye, H.; Peng, Z.; Li, R.; Chen, Y.; Wang, J.; Sun, H.; Yang, Z.; Liu, Z.; Yang, Z.; Huang, W.; Hodes, G.; Liu, S. F.; Zhao, K. Metal-Free Halide Perovskite Single Crystals with Very Long Charge Lifetimes for Efficient X-ray Imaging. *Adv. Mate.r* **2020**, *32*, e2003353.
(43) Cui, Q.; Liu, X.; Li, N.; Zeng, H.; Chu, D.; Li, H.; Song, X.; Xu, Z.; Liu, Y.; Zhu, H.; Zhao, K.; Liu, S. F. A New Metal-Free Molecular Perovskite-Related Single Crystal with Quantum Wire Structure for High-Performance X-Ray Detection. *Small* **2023**, e2308945.
(44) Xin, B.; Alaal, N.; Mitra, S.; Subahi, A.; Pak, Y.; Almalawi, D.; Alwadai, N.; Lopatin, S.; Roqan, I. S. Identifying Carrier Behavior in Ultrathin Indirect-Bandgap CsPbX(3) Nanocrystal Films for Use in UV/Visible-Blind High-Energy Detectors. *Small* **2020**, *16* (43), e2004513.
(45) Chu, D.; Jia, B.; Liu, N.; Zhang, Y.; Li, X.; Feng, J.; Pi, J.; Yang, Z.; Zhao, G.; Liu, Y.; Liu, S.; Park, N.-G. Lattice Engineering for Stabilized Black $FAPbI_3$ Perovskite Single Crystals for High-Resolution X-ray Imaging at the Lowest Dose. *Sci. Adv.* **2023**, *9*, eadh2255.

(46) Huang, Y.; Qiao, L.; Jiang, Y.; He, T.; Long, R.; Yang, F.; Wang, L.; Lei, X.; Yuan, M.; Chen, J. A-site Cation Engineering for Highly Efficient MAPbI(3) Single-crystal X-ray Detector. *Angew. Chem. Int. Ed.* **2019**, *58*, 17834-17842.

# Supporting Information for

# Machine learning aided parameter analysis in Perovskite X-ray Detector

Bobo Zhang[1†], Endai Huang[2,3†], Xinyi Du[4,5], Lu Zhang[1], Xiaokang Ma[6], Jiaxue You[7*]

[1]Key Laboratory of Applied Surface and Colloid Chemistry, Ministry of Education; Shaanxi Key Laboratory for Advanced Energy Devices; Shaanxi Engineering Lab for Advanced Energy Technology; Institute for Advanced Energy Materials; School of Materials Science and Engineering, Shaanxi Normal University, Xi'an 710119, China.

[2]Research Institute of Medical and Biological Engineering, Ningbo University, Zhejiang 315211, China

[3]Department of Computer Science and Engineering, The Chinese University of Hong Kong, Hong Kong SAR 999077, China

[4]Department of Chemical and Biomolecular Engineering, National University of Singapore, Singapore, Singapore.

[5]Solar Energy Research Institute of Singapore (SERIS), National University of Singapore, Singapore, Singapore.

[6]State Key Laboratory of Solidification Processing, Northwestern Polytechnical University, Xi'an, Shaanxi 710072, PR China.

[7]Department of Materials Science and Engineering, Hong Kong Institute for Clean Energy City University of Hong Kong, Hong Kong SAR 999077, China

†: BZ and EH contributed equally.

*Emails: jiaxuyou@cityu.edu.hk.

**Machine learning algorithm**

**Experimental details**

**Tables S1-S2**

**Figures S1-S14**

## 1. Machine learning algorithm

### 1.1. Data

We searched X-ray detectors-related papers and extracted the material features. In total, 136 samples were collected and each with 17 features, namely $Z_A$, $Z_B$, $Z_x$, $Z_{total}$, a, b, c, α, β, γ, $E_g$, μτ, RT, Trap, Trpl, Sensitivity, Detection limit. For unavailable values (e.g., features were not stated in the paper), we estimated them by experience. The prediction of lattice constants and band gaps was based on similar chemical formulas. The predictions of μτ, μ, and τ were based on the type of perovskite (thin film, crystal), similar chemical formulas, crystal dimensions (3D, 2D, 1D, 0D), and sensitivity. The prediction of resistivity was based on crystal dimensions and detection limits. The prediction of trap density was based on the type of perovskite, μ, and τ. The prediction of sensitivity was based on the type of perovskite (thin film, crystal), similar chemical formulas, crystal dimensions (3D, 2D, 1D, 0D), and μτ. The prediction of detection limits was based on crystal dimensions and similar chemical formulas.

After data collection, the data was normalized by Z-score standardization that

$$x' = \frac{x - \mu}{\sigma}$$

where $\mu$ and $\sigma$ are the average and the standard deviation of the features, respectively.

Five machine-learning algorithms for regression were selected to explore the relationship between features and objectives, namely Multivariable Linear Regression (MLR), Random Forest Regression (RFR), XGBoost, Support Vector Regression (SVR), and Genetic Programming Symbolic Regressor (GPSR). The GPSR was selected to express an explicit formula between the features and the objective. All these models were trained using the same dataset.

### 1.2. Feature importance

After the model was trained, the importance of all features was ranked from high to low. For SVR, the feature importance could not be calculated directly due to the Gaussian kernel. Therefore, a permutation importance was applied to rank the feature importance when using SVR. Permutation importance is a model-agnostic method that assesses feature importance by calculating the decrease in a model's performance when the values of each feature are randomly shuffled. This shuffling breaks the association between the feature and the target, thus the drop in model accuracy indicates the feature's predictive power. The process is repeated for each feature to obtain a ranking of their importance.

### 1.3. Weighted $R^2$ and metric of performance

To mitigate human errors when estimating the unavailable values, a sample weight was applied to estimated values that

$$W = \{w_i\}_{i=1}^{n} = \{\alpha^{a_i}\beta^{b_i}\}_{i=1}^{n}$$

where $\alpha_i$ and $\beta_i$ are the number of estimated values in features and objectives, respectively. $\alpha$ and $\beta$ are set to 0.95 and 0.9 respectively, to reduce the weight if the sample data is estimated. Thus, a weighted $R^2$ was used as a metric for the model performance that

$$R^2_{weighted} = 1 - \frac{\sum_i w_i\,(y_i - \hat{y}_i)^2}{\sum_i w_i\,(y_i - \overline{y_w})^2}$$

where $y_i$ and $\hat{y}_i$ are the true and predicted objective values, respectively, and

$$R^2_{weighted} = \frac{\sum_i w_i\,y_i}{\sum_i w_i}$$

Hold out cross-validation (HOCV) approach was applied with a training ratio of 0.8 and the average weighted R2 was used as a metric for model performance.

## 2. Experimental details

**Materials.** Hydroiodic acid (HI, ~57 wt.% in water, with 1.5% $H_3PO_2$, Aladdin), yellow lead oxide (PbO, 99.9% metals basis, orthorhombic system, Aladdin), 2-Fluorophenethylamine (m-F-PEA, 96%, Sinopharm Chemical Reagent Co., Ltd.), hypophosphorous acid solution ($H_3PO_2$, 50 wt. %, Aladdin). All reagents are used directly without further processing.

**Growth of (m-F-PEA)$_2$PbI$_4$ single crystals (SCs).** Single crystal (m-F-PEA)$_2$PbI$_4$ was synthesized by the temperature cooling crystallization (TCC) method. First, PbO (5 mmol) and o-F-PEAI (10 mmol) were dissolved in a mixture of 30 ml HI and 3.75 ml $H_3PO_2$ and heated to 105 °C, producing a clear-yellow solution. To ensure proper dissolution and complete reaction, the solution was stirred vigorously for 24 hours at 105 °C. The solution was then filtered into a clear crystallizing dish using a 0.45-μm pore-size mixed-cellulose water system filter head. The crystallizing dish containing the crystal growth solution was sealed with plastic wrap and quickly transferred into an oven at 105 °C constant temperature for 10h and then cooled slowly to 40 °C at a rate of 5 °C per day. Finally, centimeter-sized (m-F-PEA)$_2$PbI$_4$ SCs were obtained.

**Device Fabrication.** The (m-F-PEA)$_2$PbI$_4$ SC X-ray detector was made by depositing interdigital Au electrodes (~60 nm thickness) with a coplane structure. The effective area of the device was $1.2 \times 10^{-3}$ cm$^2$. The interdigital electrode distance is 40 μm.

**Characterization of the materials.**

**X-ray diffraction (XRD).** The powder XRD analysis was conducted using a Rigaku (Smartlab-9kW) X-ray diffractometer, which was equipped with a Cu Kα X-ray ($\lambda$=1.54186 Å) tube. The instrument was operated at 40 kV and 30 mA, with a scanning step of approximately $0.02° \cdot s^{-1}$.

**Steady-state photoluminescence (PL).** The investigation of the PL in single crystals was carried out employing a Pico-Quant FT-100 spectrometer, with the excitation light source being a laser operating at a wavelength of 510 nm.

**Thermogravimetric analysis (TGA).** In order to determine the TGA curve, meticulous grinding was performed on a sample of SC to obtain a fine powder. The TGA was conducted utilizing the TA SDT-Q600 V20.9 (Build 20) instrument. Approximately 5 mg of the SC powder was carefully placed in an $Al_2O_3$ crucible. Subsequently, the temperature was gradually raised to 800 °C at a controlled rate of 10 °C $min^{-1}$, all while maintaining a nitrogen environment with a gas flow rate of 100 mL $min^{-1}$.

**Scanning electron microscope (SEM).** The surface and cross-sectional morphologies of the crystal were examined under a SEM model SU-8020, manufactured by Hitachi. The observations were conducted in a controlled vacuum environment to ensure accurate characterization.

**UV-Vis-NIR absorbance spectroscopy.** The acquisition of the UV-Vis-NIR absorbance spectrum was accomplished through the utilization of a Perkin-Elmer Lambda 950 UV-Vis-NIR spectrophotometer. The spectral measurement spanned from 300 to 800 nm, and this particular analysis was executed under ambient temperature conditions.

**Band structure calculation**

We have employed the first-principles to perform by density functional theory (DFT) calculations have been carried out by using Vienna ab initio Simulation Package (VASP)[1]. The exchange correlation potential is described by a generalized gradient approximation (GGA) with a Perdew-Burke-Ernzerhof (PBE) functional and the cut-off energy of 500 eV was used due to the plane wave expansion[2]. For structural relaxation, the convergence criteria for energy and force are $10^{-5}$ eV/atom and 0.05 eV/Å, respectively. $6 \times 10 \times 4$ Monkhorst-Pack k-mesh Brillouin zone sampling scheme was used for geometric optimization.

**Dielectric constant analysis.** An experimental setup involving a simple plate capacitor was employed to determine the dielectric constant of $(m\text{-}F\text{-}PEA)_2PbI_4$ SCs. The device consisted of two layers of gold (Au) electrodes, with the $(m\text{-}F\text{-}PEA)_2PbI_4$ SCs acting as the dielectric material in between. The measurement of capacitance was conducted as a function of frequency under dark conditions. The relative dielectric constant of $(m\text{-}F\text{-}PEA)_2PbI_4$ was then calculated using the parallel plate capacitor equation:[3]

$$C = \frac{\varepsilon_0 \varepsilon S_a}{D_t}$$

where $S_a$ is the electrode area (1.77 $mm^2$), and $D_t$ is the thickness of the SC.

**Space-charge-limited current (SCLC) measurement.** The widely accepted space charge-limited current (SCLC) method was utilized to determine the trap density ($n_{trap}$) of the $(m\text{-}F\text{-}PEA)_2PbI_4$ SCs. The hole-device configuration consisted of a 60 nm thick Au layer sandwiched between two layers of $(m\text{-}F\text{-}PEA)_2PbI_4$, with Au serving as the electrodes. The electrode area was measured to be 1.77 $mm^2$, while the crystal thickness was 0.6 mm. A Keysight B2902A source meter was employed to measure the dark I-V curve, which was then analyzed based on the SCLC model. The voltage corresponding to the transition point from Ohmic behavior to the trap-filled region is commonly referred to as the trap-filled-limit voltage ($V_{TFL}$). Thus, the trap density can be calculated using the provided equation:[4]

$$n_{\text{trap}} = \frac{2\varepsilon_0 \varepsilon V_{TFL}}{qL^2}$$

where $\varepsilon_0$ is the vacuum dielectric constant, $\varepsilon$ the relative dielectric constant, $L$ the thickness of the sample.

**Photoconductivity analysis.** An analysis of photoconductivity was conducted on the vertical device consisting of $Au/(m\text{-}F\text{-}PEA)_2PbI_4/Au$. The current-voltage curve was fitted using a modified version of the Hecht equation,[5] resulting in the determination of the product of mobility and carrier lifetime (μτ) and the surface recombination velocity (S):

$$I = \frac{I_0 \mu \tau V}{L^2} \frac{1 - \exp\left(-\frac{L^2}{\mu \tau V}\right)}{1 + \frac{L}{V}\frac{S}{\mu}}$$

where $I_0$ is the saturated photocurrent, $L$ is the thickness, $V$ is the applied bias, and $\tau$ is the carrier lifetime.

**Dark current drift measurements.** The measurement of dark current drift involves recording the current of the devices in a dark environment while maintaining a constant electric field. The calculation of dark current drift is based on the data provided in the published report as:[6]

$$I_{\text{draft}} = \frac{I_{finish} - I_{begin}}{t \cdot s \cdot E}$$

where $I_{begin}$ and $I_{finish}$ are the recorded initial and final dark current after a certain measure time (*t*), 12960 s, *s* is the distance between the two interdigital electrodes, $5.0\times10^{-3}$ cm, *E* the applied electric field, 10 V, respectively.

**X-ray detector performance measurements.** The X-ray detection performance was evaluated within a custom-made dark box to minimize any interference from electromagnetic and ambient light. A tungsten anode X-ray tube (DX-DS2901/24) served as the source of X-rays. The X-ray tube operated at a constant 40 kV voltage, while the current was adjusted from 40 mA to 5 mA to control the dose rate of the X-rays. To precisely adjust the dose rate, 2-mm-thick Al foils were utilized as attenuators. The X-ray dose rate was carefully calibrated using a Fluke Si diode (RaySafe X2 R/F) dosimeter. Throughout the measurements, the X-ray response current was recorded using a high-precision source meter (Keysight B2902A). All measurements were conducted at room temperature within the scanning system's window.

**Table S1**. Comparison of prediction accuracy ($R^2$) of four models for four targets.

| Model | Dataset | $E_g$ | $\mu\tau$ | Sensitivity | Detection |
|---|---|---|---|---|---|
| MLR | Train | 0.26 | -0.09 | -1.13 | -0.10 |
| | Test | 0.81 | 0.26 | 0.08 | 0.03 |
| RFR | Train | 0.93 | 0.84 | 0.53 | 0.80 |
| | Test | 0.80 | 0.17 | 0.08 | 0.07 |
| XGBoost | Train | 0.99 | 0.83 | 0.32 | 0.48 |
| | Test | 0.72 | 0.16 | 0.02 | 0.07 |
| SVR | Train | 0.97 | 0.29 | 0.07 | 0.16 |
| | Test | 0.72 | 0.08 | 0.03 | 0.02 |
| GPSR | Train | 0.72 | -0.16 | 0.15 | -0.15 |
| | Test | 0.61 | -0.19 | -0.34 | -0.16 |

**Table S2**. Crystal data and structure refinement for the $(m\text{-}F\text{-}PEA)_2PbI_4$ crystal.

| | |
|---|---|
| Empirical formula | $C_{16}H_{22}F_2N_2PbI_4$ |
| Crystal system | monoclinic |
| Temperature | 298 K |
| Space group | P2/c |
| Unit cell dimensions | a =66.489(4) Å, α=90°<br>b =12.2142(6) Å, β=95.086(2)°<br>c =12.2131(9) Å, γ=90° |
| Volume | 9879.3(10) $Å^3$ |
| Density (calculated) | 2.676 (g $cm^{-3}$) |
| Z | 4 |
| μ | 11.849 $mm^{-1}$ |
| F(000) | 7104.0 |
| 2Θ range for data collection | 3.69 to 50.7° |
| Index ranges | $-80 \leq h \leq 79$, $-13 \leq k \leq 14$, $-14 \leq l \leq 14$ |
| Reflections collected | 33606 |
| Independent reflections | 9016 [$R_{int}$ = 0.0636, $R_{sigma}$ = 0.0622] |
| Data/restraints/parameters | 9016/503/542 |
| Goodness-of-fit on $F^2$ | 1.078 |
| Final R indexes [I>=2σ (I)] | $R_1$ = 0.0987, $wR_2$ = 0.2541 |
| Final R indexes [all data] | $R_1$ = 0.1246, $wR_2$ = 0.2919 |
| Largest diff. peak/hole | 4.65/-7.15 $Å^{-3}$ |

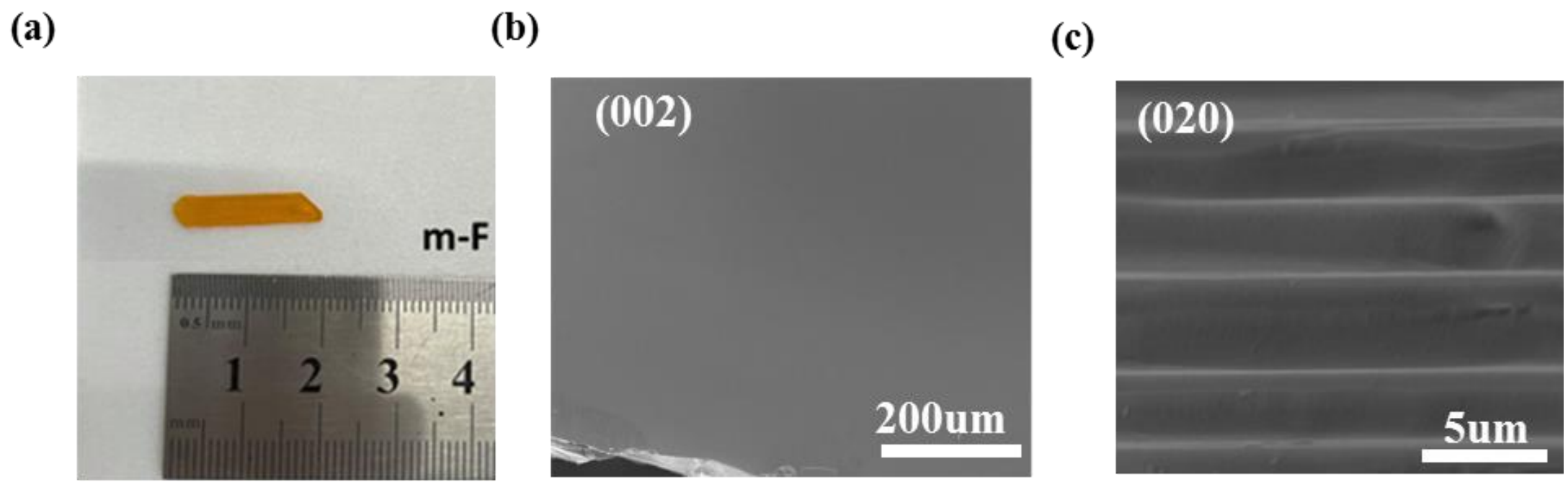

**Figure S1.** (a) Photograph of an $(m\text{-}F\text{-}PEA)_2PbI_4$ single crystal. (b) Top-view and (c) cross-sectional SEM images of $(m\text{-}F\text{-}PEA)_2PbI_4$ single crystal.

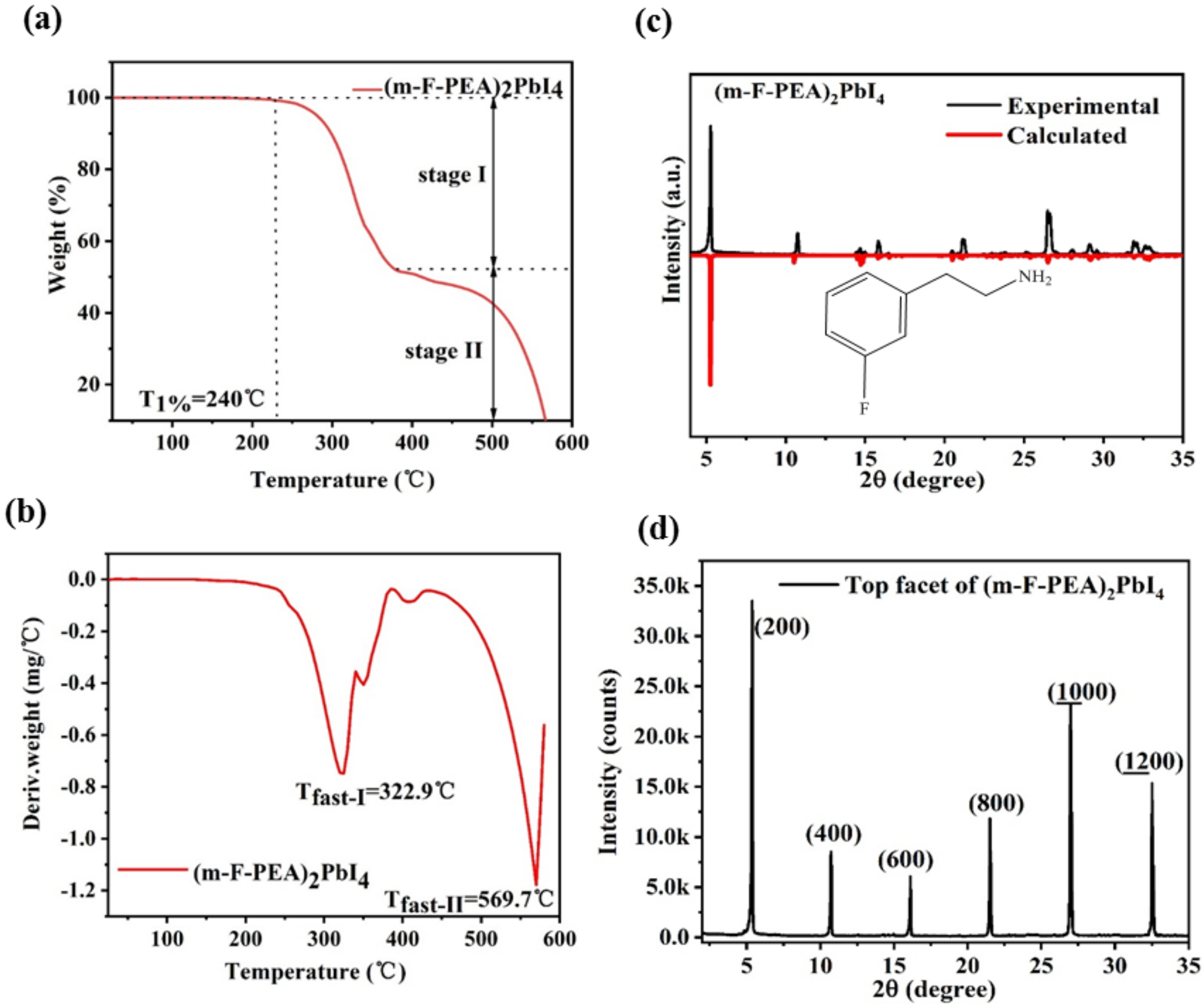

**Figure S2.** (a) TGA curve of the $(m\text{-}F\text{-}PEA)_2PbI_4$ single crystal and (b) the corresponding first derivative. (c) Comparison of the measured powder XRD and the calculated XRD patterns of the $(m\text{-}F\text{-}PEA)_2PbI_4$. (d) XRD pattern of the $(m\text{-}F\text{-}PEA)_2PbI_4$ single crystal.

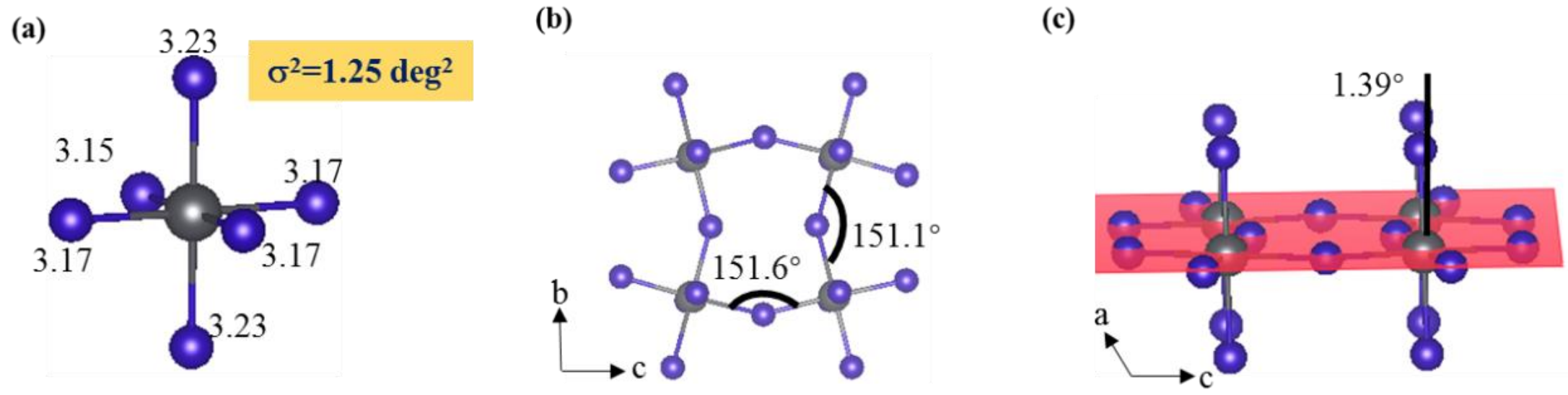

**Figure S3.** (a) Ball-and-stick model of lead iodine octahedron. (b) In-plane (Pb-I-Pb) and (c) out-of-plane distortion (quantitatively expressed as the residual angle of the angle formed by the Pb-I bond in the stacking direction and the layer plane) of (m-F-PEA)$_2$PbI$_4$.

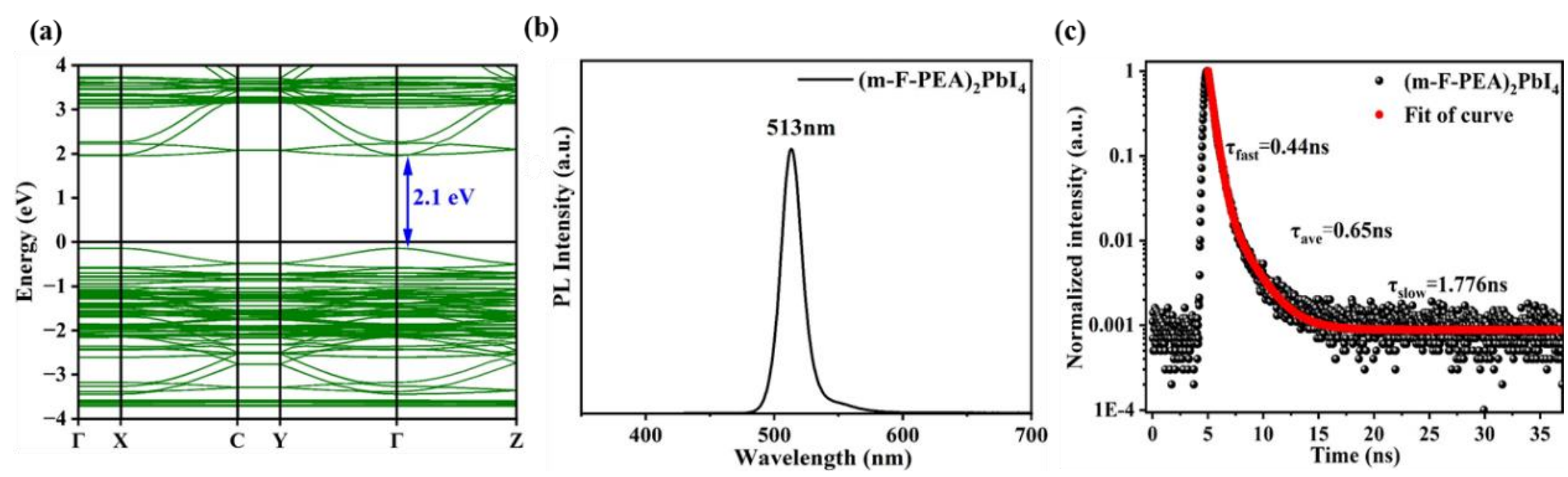

**Figure S4.** (a) band structure, (b) PL spectra and (c) the corresponding PL lifetime for the (m-F-PEA)$_2$PbI$_4$.

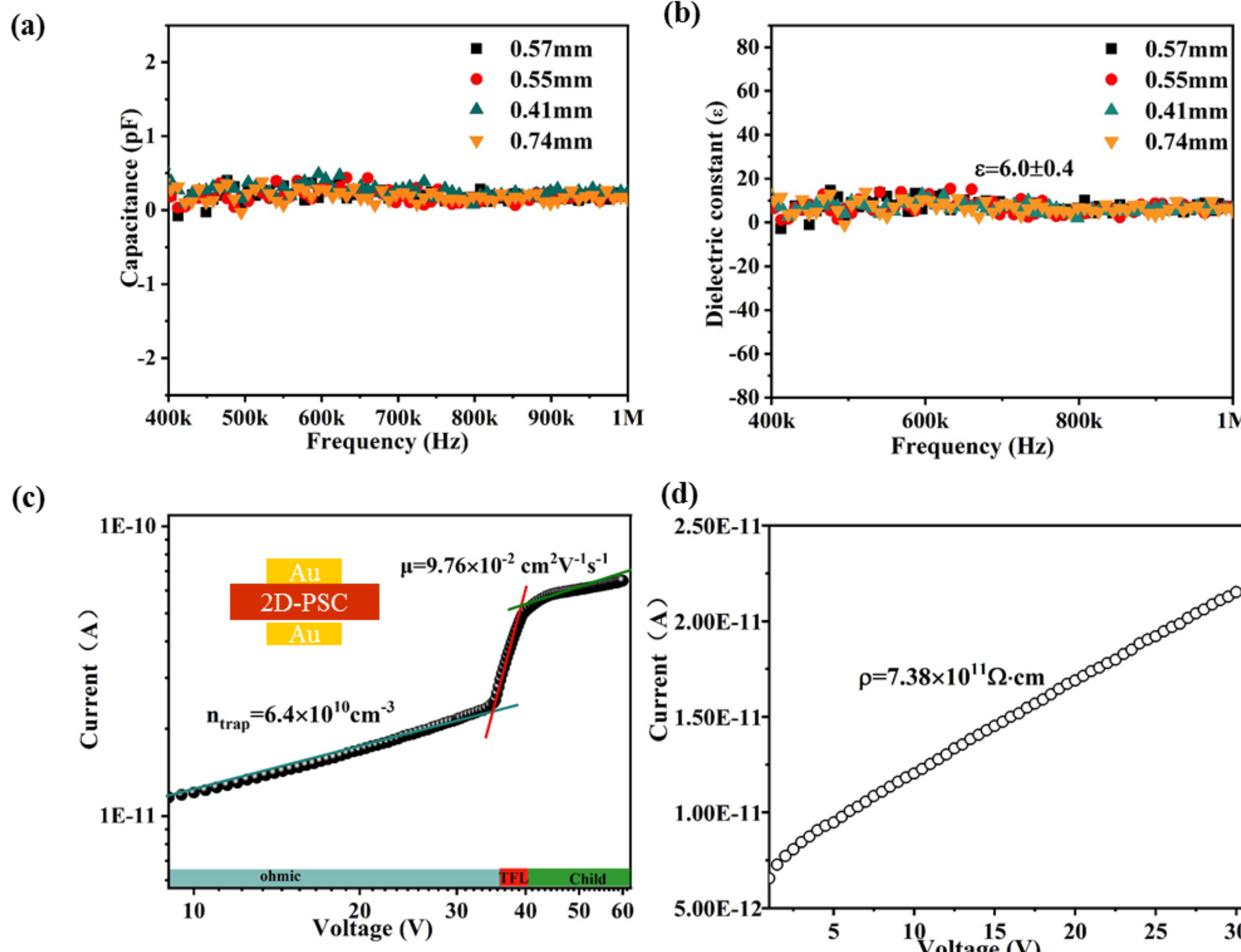

**Figure S5.** (a) Frequency-dependent capacitance of the (m-F-PEA)$_2$PbI$_4$ SCs. (b) Dielectric constants of the (m-F-PEA)$_2$PbI$_4$ SCs. (c) Dark I−V curves with determined trap-filled limit voltage of the (m-F-PEA)$_2$PbI$_4$ SC. Insets: structure of the devices used for measurements. (d) Bulk resistivity of (m-F-PEA)$_2$PbI$_4$ SC.

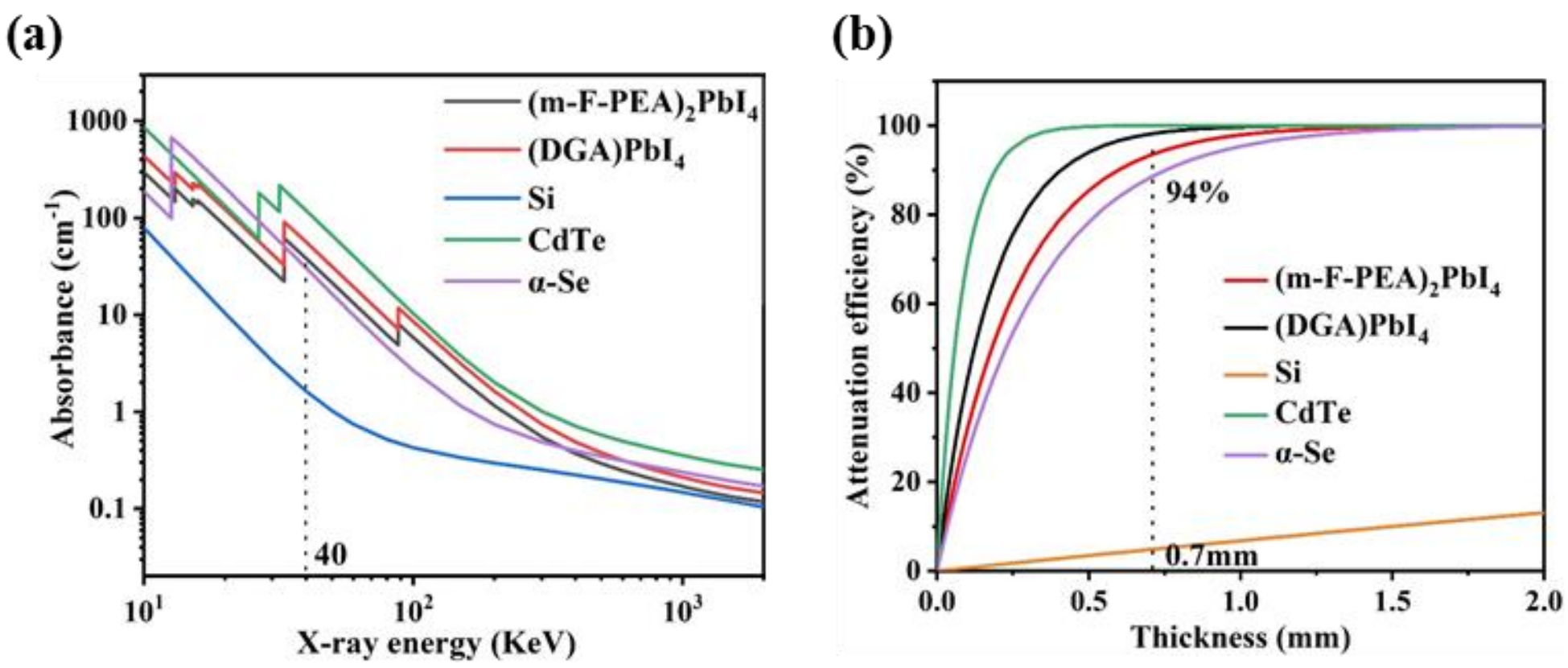

**Figure S6.** (a) Absorption coefficients of the (m-F-PEA)$_2$PbI$_4$, (DGA)PbI$_4$, CdZnTe, α-Se and Si to the X-rays with different photon energy. (b) Attenuation efficiency of the (m-F-PEA)$_2$PbI$_4$, (DGA)PbI$_4$, CdZnTe, α-Se and Si with different thickness for the 40 keV X-ray photons.

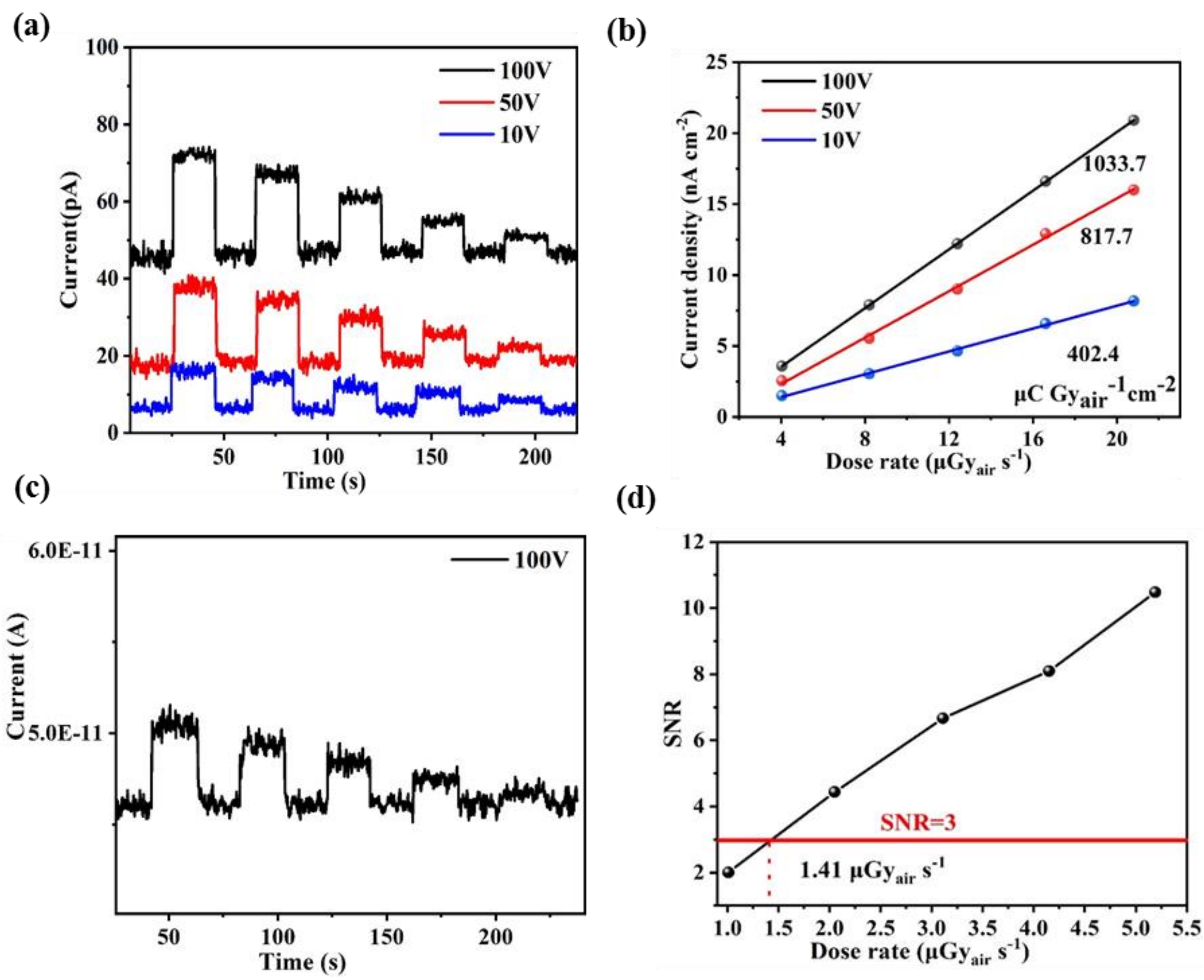

**Figure S7.** (a) The I-T curve of the device at different doses varies from 10-100V with the bias voltage. (b) Sensitivity fitting of the (m-F-PEA)$_2$PbI$_4$ SC detector under different bias voltage. (c) I-T curve of (m-F-PEA)$_2$PbI$_4$ SC device at 100V bias and low dose (5.19, 4.15, 3.11, 2.05, 1.01 $\mu Gy_{air}$ $s^{-1}$). (d) Signal to noise ratio (SNR) of the (m-F-PEA)$_2$PbI$_4$ SC detector under different dose rates at 100 V bias.

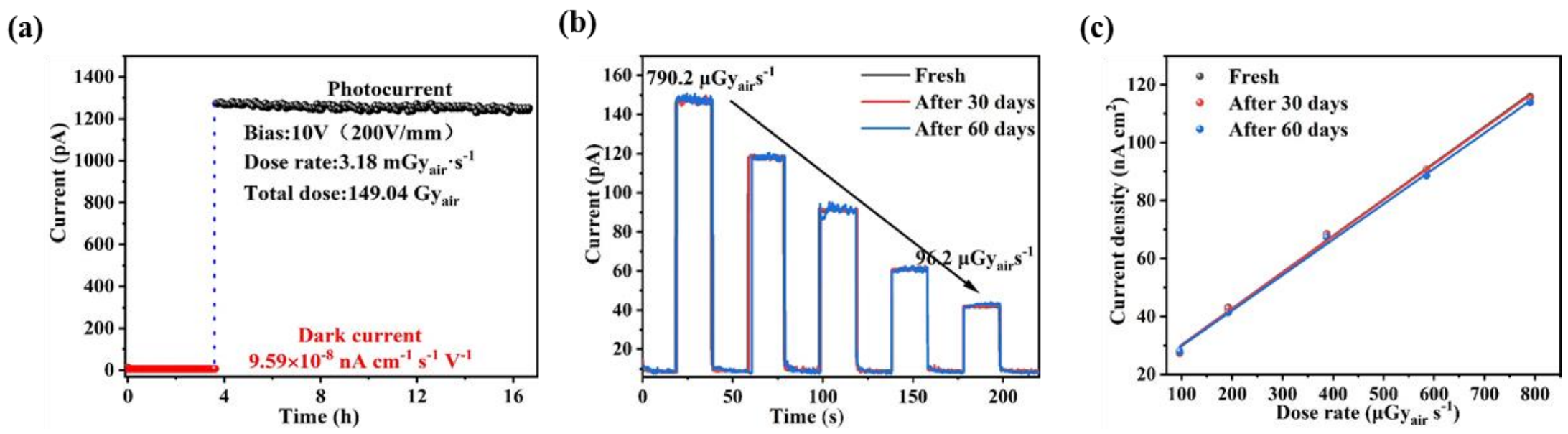

**Figure S8.** (a) Dark current drift and radiation stability (under continuous X-ray irradiation for more than 12 h) measurement of the $(m\text{-}F\text{-}PEA)_2PbI_4$ SC detector. (b) I-T response curves to X-rays for fresh devices and devices after exposure to air for 30 days and 60 days, respectively. (c) X-ray photocurrent densities at different dose rates under fresh conditions and after 30 days, 60 days ambient air ageing, respectively.

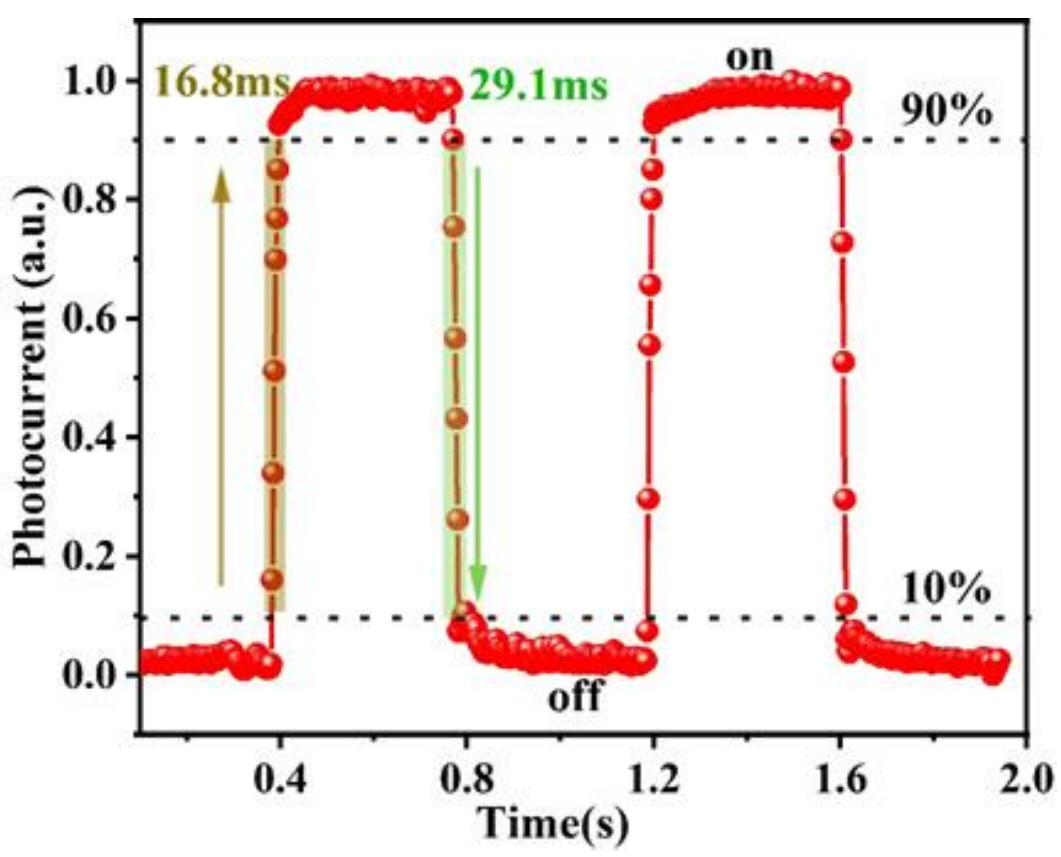

**Figure S9.** Transient response of the $(m\text{-}F\text{-}PEA)_2PbI_4$ SC detector to pulsed X-rays at a 10V bias voltage.

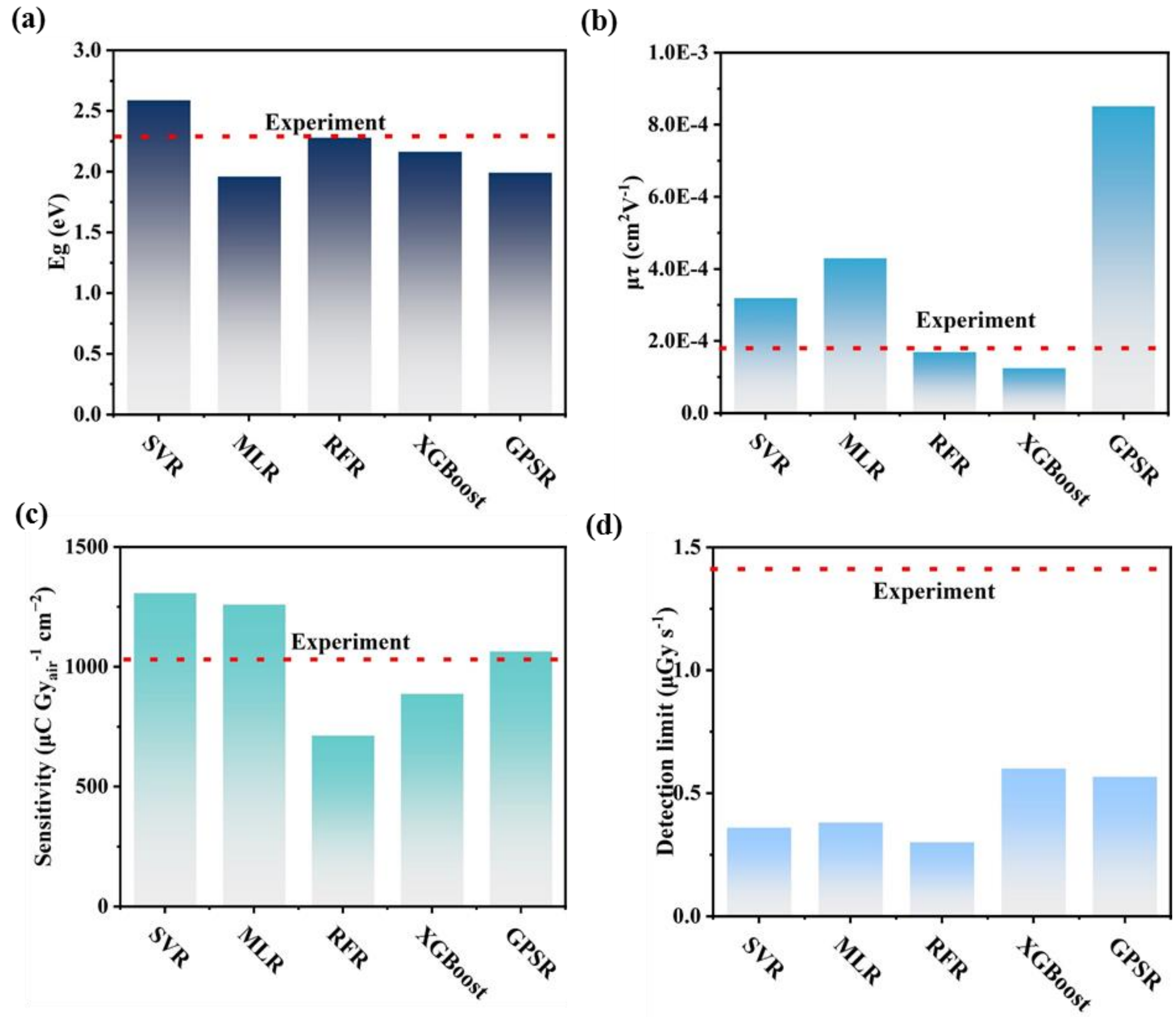

**Figure S10.** Comparison of experimental values ($E_g$, μτ, sensitivity, detection limit) of $(m\text{-}F\text{-}PEA)_2PbI_4$ with the predicted values of these four parameters by five different ML models.

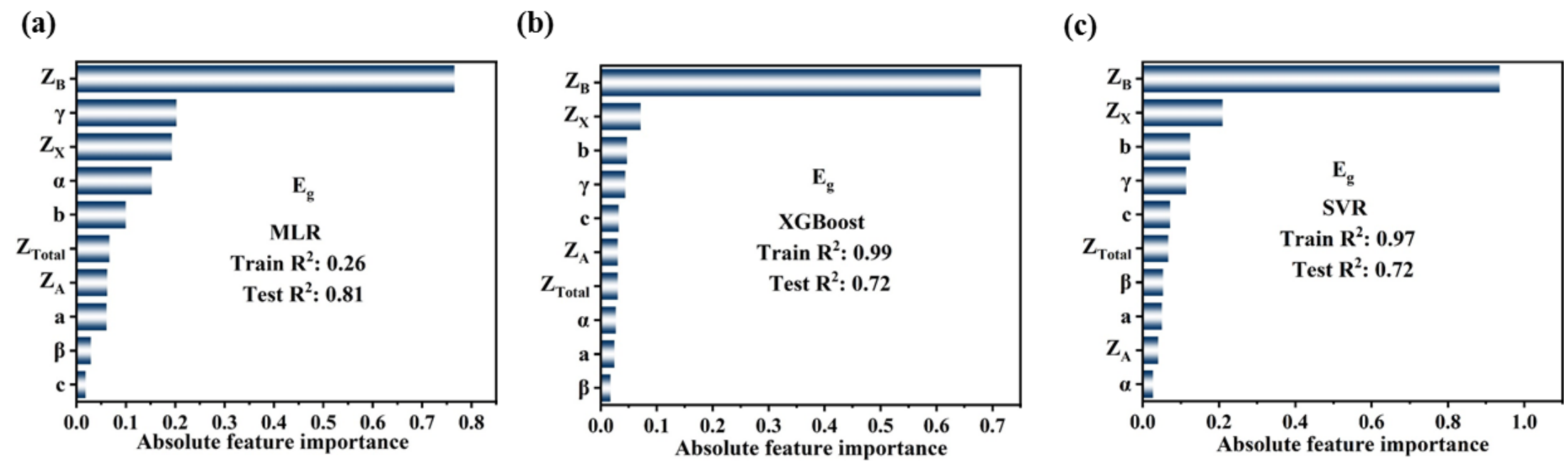

**Figure S11.** (a-c) Feature importance ranking affecting $E_g$ under different algorithms. The weight of the GPSR is not shown, as it outputs a constant close to the average of the target, indicating that no relationship is captured.

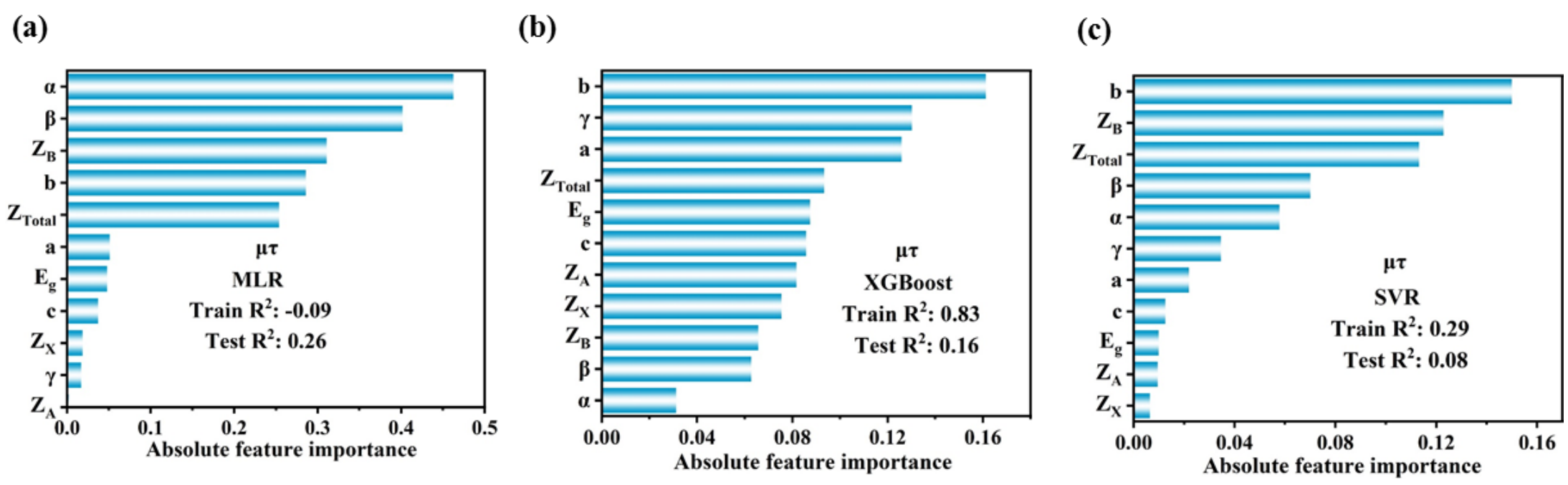

**Figure S12.** (a-c) Feature importance ranking affecting μτ under different algorithms. The weight of the GPSR is not shown, as it outputs a constant close to the average of the target, indicating that no relationship is captured.

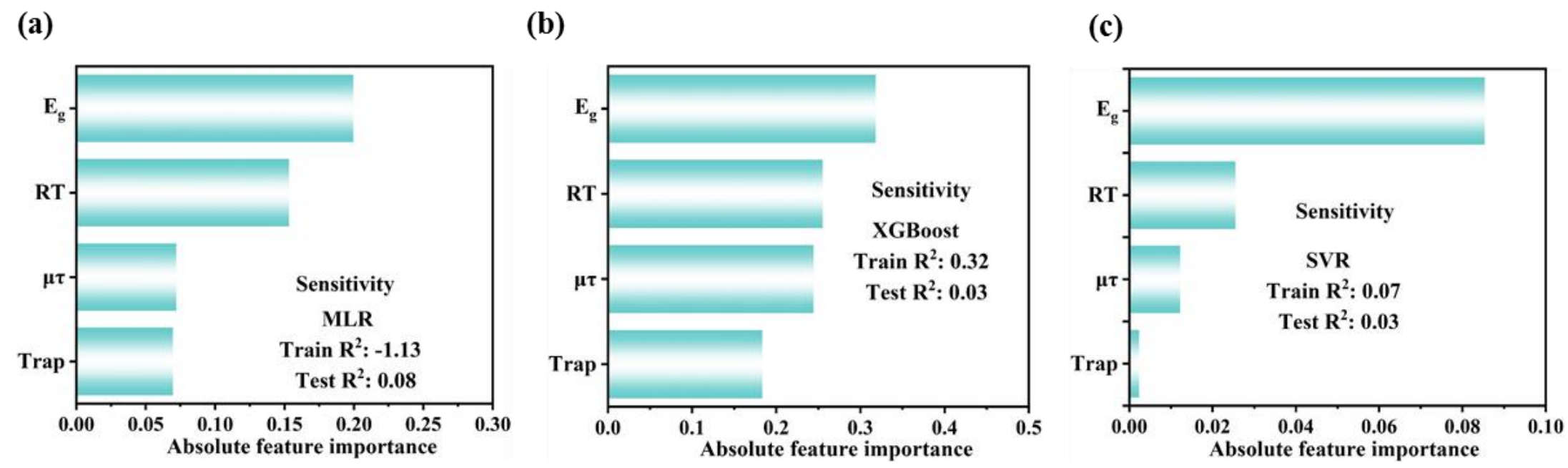

**Figure S13.** (a-c) Feature importance ranking affecting sensitivity under different algorithms. The weight of the GPSR is not shown, as it outputs a constant close to the average of the target, indicating that no relationship is captured.

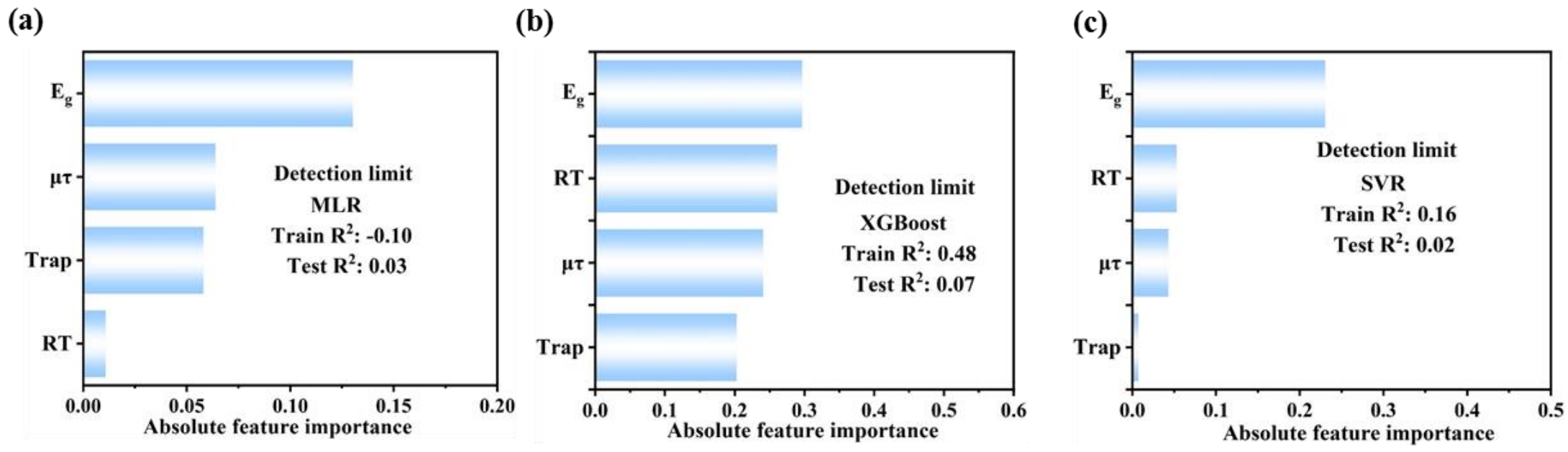

**Figure S14.** (a-c) Feature importance ranking affecting detection limit under different algorithms. The weight of the GPSR is not shown, as it outputs a constant close to the average of the target, indicating that no relationship is captured.

## References

(1) Kresse, G.; Joubert, D. From Ultrasoft Pseudopotentials to the Projector Augmented-wave Method. *Phys. Rev. B* **1999**, *59*, 1758-1775.

(2) Perdew, J. P.; Burke, K.; Ernzerhof, M. Generalized Gradient Approximation Made Simple. *Phys. Rev. Lett.* **1996**, *77*, 3865-3868.

(3) Li, K.; Dong, L.-Y.; Xu, H.-X.; Qin, Y.; Li, Z.-G.; Azeem, M.; Li, W.; Bu, X.-H. Electronic Structures and Elastic Properties of a Family of Metal-free Perovskites. *Mater. Chem. Front.* **2019**, *3*, 1678-1685.

(4) Pan, W.; Wu, H.; Luo, J.; Deng, Z.; Ge, C.; Chen, C.; Jiang, X.; Yin, W.-J.; Niu, G.; Zhu, L.; Yin, L.; Zhou, Y.; Xie, Q.; Ke, X.; Sui, M.; Tang, J. $Cs_2AgBiBr_6$ Single-crystal X-ray Detectors with a Low Detection Limit. *Nat. Photonics* **2017**, *11*, 726-732.

(5) Androulakis, J.; Peter, S. C.; Li, H.; Malliakas, C. D.; Peters, J. A.; Liu, Z.; Wessels, B. W.; Song, J. H.; Jin, H.; Freeman, A. J. Dimensional Reduction: A Design Tool for New Radiation Detection Materials. *Adv. Mater.* **2011**, *23*, 4163-4167.

(6) Yang, B.; Pan, W.; Wu, H.; Niu, G.; Yuan, J.-H.; Xue, K.-H.; Yin, L.; Du, X.; Miao, X.-S.; Yang, X.; Xie, Q.; Tang, J. Heteroepitaxial Passivation of $Cs_2AgBiBr_6$ Wafers with Suppressed Ionic Migration for X-ray Imaging. *Nat. Commun.* **2019**, *10*, 1989.